\documentclass[twocolumn,aps,prl,floatfix,superscriptaddress]{revtex4-1}

\usepackage{amsmath,amssymb}
\usepackage{graphicx,float}
\usepackage{times,txfonts}
\usepackage{color}
\usepackage{multirow}
\usepackage{bbold}
\usepackage{color}
\usepackage{soul}
\usepackage{hyperref}
\usepackage{times,txfonts}
\usepackage{natbib}
\usepackage{nicefrac}
\usepackage{blindtext}
\usepackage[export]{adjustbox}
\usepackage{mathrsfs}
\usepackage{textcomp}
\usepackage{multirow}
\usepackage[normalem]{ulem}
\usepackage{amsmath}

\newcommand{\abs}[1]{\left| #1 \right|}

\newcommand{\bra}[1]{\left\langle #1 \right|}
\newcommand{\ket}[1]{\left| #1 \right\rangle}
\newcommand{\braket}[2]{\left\langle {#1{\left| \vphantom{#1 #2} \right.} #2} \right\rangle}
\newcommand{\qo}[1]{``#1''}

\renewcommand{\epsilon}{\varepsilon}

\def\VR{\kern-\arraycolsep\strut\vrule &\kern-\arraycolsep}
\def\vr{\kern-\arraycolsep & \kern-\arraycolsep}
\usepackage{soul}
\definecolor{lightblue}{RGB}{185,210,248}
\definecolor{lgreen}{RGB}{15,150,15}
\sethlcolor{lightblue}

\begin{document}

\title{Quantum tomography of inelastic electron scattering \emph{via} orbital angular momentum states}

\author{Amir~H.~Tavabi}
\affiliation{Ernst Ruska-Centre for Microscopy and Spectroscopy with Electrons, Forschungszentrum Jülich, 52425 Jülich, Germany}

\author{Alessio~D'Errico}
\affiliation{Nexus for Quantum Technologies, University of Ottawa, Ottawa ON, Canada, K1N 5N6}

\author{Paolo~Rosi}
\affiliation{Istituto Nanoscienze, Consiglio Nazionale delle Ricerche, Via G. Campi 213/A, 41125 Modena, Italy}

\author{Giovanni~Bertoni}
\affiliation{Istituto Nanoscienze, Consiglio Nazionale delle Ricerche, Via G. Campi 213/A, 41125 Modena, Italy}

\author{Enzo~Rotunno}
\affiliation{Istituto Nanoscienze, Consiglio Nazionale delle Ricerche, Via G. Campi 213/A, 41125 Modena, Italy}

\author{Luca~Belsito}
\affiliation{Istituto per la Microelettronica e i Microsistemi, Consiglio Nazionale delle Ricerche, Via P. Gobetti 101, 40129 Bologna, Italy}

\author{Alberto~Roncaglia}
\affiliation{Istituto per la Microelettronica e i Microsistemi, Consiglio Nazionale delle Ricerche, Via P. Gobetti 101, 40129 Bologna, Italy}

\author{Stefano~Frabboni}
\affiliation{Istituto Nanoscienze, Consiglio Nazionale delle Ricerche, Via G. Campi 213/A, 41125 Modena, Italy}
\affiliation{Università di Modena e Reggio Emilia, Via G. Campi 213/A, 41125 Modena, Italy}

\author{Gian~Carlo~Gazzadi}
\affiliation{Istituto Nanoscienze, Consiglio Nazionale delle Ricerche, Via G. Campi 213/A, 41125 Modena, Italy}

\author{Peter~Tiemeijer}
\affiliation{Thermo Fisher Scientific, PO Box 80066, 5600 KA Eindhoven, the Netherlands}

\author{Rafal~E.~Dunin-Borkowski}
\affiliation{Ernst Ruska-Centre for Microscopy and Spectroscopy with Electrons, Forschungszentrum Jülich, 52425 Jülich, Germany}

\author{Ebrahim~Karimi}
\email{ekarimi@uottawa.ca}
\affiliation{Nexus for Quantum Technologies, University of Ottawa, Ottawa ON, Canada, K1N 5N6}
\affiliation{National Research Council of Canada, 100 Sussex Drive, Ottawa ON, Canada, K1A 0R6}
\affiliation{Institute for Quantum Studies, Chapman University, Orange, California 92866, USA}

\author{Vincenzo~Grillo}
\affiliation{Istituto Nanoscienze, Consiglio Nazionale delle Ricerche, Via G. Campi 213/A, 41125 Modena, Italy}

\begin{abstract}
The physical properties of a quantum system, whether pure or mixed, are described fully by its density matrix. Recovery of the density matrix through projective measurements -- referred to as quantum state tomography -- is a cornerstone of quantum optics and metrology. The implementation of this approach in transmission electron microscopy, in particular for the characterisation of an electron beam after inelastic scattering, has remained a longstanding challenge as a result of the complexity of scanning high-dimensional phase spaces, with the number of required measurements growing quadratically with space dimensionality. Here, we introduce a simplified approach by restricting tomography to the electron orbital angular momentum (OAM) subspace. By using an electron optical device known as an OAM sorter, we discretise the phase space into a finite set of measurable states, thus significantly reducing the experimental and computational burden. The resulting measurements suffice to probe essential features of inelastic scattering. We demonstrate the technique by studying the inelastic scattering of a structured electron probe exciting volume plasmons in a carbon film. The combined use of a structured beams and OAM-resolved quantum tomography reveals symmetry-breaking effects and offers insight into the coherence and evolution of the scattered quantum states. Analysis of the diagonalised density matrices further reveals the nature of the induced state transitions, demonstrating the power of the approach for quantum tomography of electron scattering.
\end{abstract}	

\maketitle

\section{introduction}
The concepts of a density matrix and, subsequently, quasiprobability functions (\emph{e.g.}, Wigner, Glauber-Sudarshan (P) and Husimi (Q)), as representations of quantum states were developed within the frameworks of quantum statistical mechanics and the theory of quantum measurements~\cite{Cohen-Tannoudji1991}. These formalisms allow for a rigorous description of both pure and mixed quantum states and provide powerful tools for quantifying a quantum system’s physical features and the information that is exchanged during interactions with other systems or \emph{environments}. They also enable the identification of entanglement signatures and appropriate \qo{witnesses} for nonclassical correlations in complex quantum systems. Notably, the loss of coherence of a probe --manifested in a given basis -- can, by itself, carry valuable indirect information about the nature of the interacting system. In the special case of purely elastic scattering with an initial pure-state probe (\emph{i.e.}, negligible decoherence), quantum tomography simplifies to a phase retrieval problem~\cite{lundeen2011direct,liberman2016quantum,PhysRevLett.105.150401,compressivesensing,dehghan2024biphoton}. In such a scenario, the state remains coherent, and the task reduces to reconstructing the phase of the wavefunction rather than its full density matrix. However, inelastic scattering fundamentally changes the problem, as the introduction of mutual incoherence between the scattered components results in a mixed quantum state, significantly increasing the complexity and dimensionality of the inverse problem~\cite{schattschneider1999density,schattschneider2000physical}. Although pure states can be described using a simple wavefunction formalism, mixed states demand a full density matrix representation, as the wavefunction approach inherently assumes coherent superpositions -- a condition that no longer holds.
In electron microscopy, various inverse methods have been developed to recover the phase information of the electron beam. Many of these methods, such as multi-plane phase retrieval based on the Gerchberg-Saxton algorithm and, more generally, error reduction techniques~\cite{Gerchberg1972,Fienup1982}, rely on intensity measurements in different planes to infer missing phase information. An instructive example is provided by electron ptychography \cite{rodenburg1992ptychography,Jiang2018}, where a scanning coherent probe explores both position and momentum space, two conjugate planes, thereby ensuring sufficient sampling to enable reliable phase reconstruction. Certain ptychographic algorithms leverage the fact that the recorded data can be interpreted as a convolution of the probe’s and the sample’s Wigner functions -- a phase space reformulation of the density matrix -- under the assumption that both the probe and the sample remain in pure states~\cite{Yang2017}. This strong constraint enables well-posed inversion. Extension of these techniques to accommodate mixed-state probes is an active area of research. Iterative phase retrieval algorithms, in particular those based on error reduction methods, are more naturally suited to generalisation toward mixed states. In this context, inelastic scattering ptychography has emerged as a sought-after goal in the electron microscopy community, potentially unveiling richer structural and dynamical information that is inaccessible using elastic imaging alone. Early demonstrations of quantum tomography in electron microscopy were necessarily simplified, often introducing an envelope function to model spatial decoherence effects~\cite{Roder2014}. However, such an approach suffers from limitations when applied to genuine inelastic scattering events, in which more complex interactions blur the assumptions that underpin the envelope function.
A more consistent realisation of full quantum state tomography in electron microscopy has so far been achieved not in conventional transmission electron microscopes (TEMs) but within the framework of ultrafast transmission electron microscopy (UTEM), when electrons interact coherently with optical fields and the flexibility to control and manipulate the light state enables precise reconstruction of the electron’s quantum state~\cite{priebe2017attosecond}. Techniques such as Ramsey-type holography in UTEM represent another avenue for reconstructing inelastic phase evolution, although they typically stop short of realising full quantum state analysis~\cite{bucher2023free,gaida2024attosecond}.
Meanwhile, methods such as orbital mapping~\cite{iwashimizu2021electron} and centre-of-mass mapping~\cite{haas2022advances}, which measure a the  spatial distribution of a fine electron probe or momentum-resolved inelastic scattering cross-sections, provide only partial access to the quantum state. These methods typically capture diagonal elements -- and occasionally limited phase information -- of the density matrix projections, but reveal little about the mutual coherences that are essential to achieve full quantum state characterisation.

Here, we apply the recently-developed orbital angular momentum (OAM) sorter~\cite{Grillo2017,Tavabi2021sorter} to constrain measurements to the angular degree of freedom, effectively tracing over the radial degree of freedom. Within this basis, the relevant conjugate variables are azimuthal angle and OAM, which are related to each other through a Fourier transform.

\section{Results}
\noindent{\textbf{OAM space and quantum state tomography:}} 
The OAM of an electron is intrinsically linked to the helical phase structure of its quantum wavefunction $\ket{\psi}$. In cylindrical coordinates $(r, \theta, z)$, the wavefunction of an OAM eigenstate propagating along the $z$-axis can be expressed as $\psi(\mathbf{r}) \propto e^{i m \theta}$, where $m$ is an integer known as the \emph{topological charge}, corresponding to a quantised value of OAM in units of $\hbar$~\cite{bliokh:2023}. In this representation, the radial dependence, which can be characterised separately by a radial index, has been neglected. The radial degree of freedom may be either continuous or discrete, depending on the type of spatial mode considered~\cite{karimi:2012,karimi:2014a, karimi:2014b,plick:2015}. For instance, it is continuous for Bessel-Gauss beams and discrete for Laguerre-Gauss modes. As a result of the inherent quantisation of OAM, the electron’s quantum state space in the OAM degree of freedom is spanned by a discrete set of eigenstates $\{ \ket{m} \}$. Consequently, the density matrix $\hat{\rho}=\sum_{m,m'}\rho_{m,m'}\,\ket{m}\!\bra{m'}$ representing a general, possibly mixed, quantum state in this subspace adopts a finite, compact (parsimonious) matrix representation. Each element $\rho_{m,m'}=\bra{m}\rho\ket{m'}$ of the density matrix encodes the coherence between the OAM modes $\ket{m}$ and $\ket{m'}$, with the diagonal terms $\rho_{m,m}=\bra{m}\rho\ket{m}$ corresponding to the probabilities associated with each OAM state. This discrete structure simplifies both the theoretical formulation and the practical reconstruction of the quantum state, reducing the problem from estimating a continuous function over phase space to determining a finite number of matrix elements. Here, we exploit this property to perform quantum tomography restricted to the OAM subspace, on the assumption that contributions from different radial modes are either negligible or effectively traced out.

In order to fully reconstruct the density matrix $\hat{\rho}$ in the OAM subspace, it is necessary to perform a \emph{complete} set of projective measurements, whose number and structure depend on the dimension $d$ of the Hilbert space. If the accessible OAM values are restricted to the set $\{ -\ell, \ldots, +\ell \}$, then the dimension of the corresponding space is $d = 2\ell + 1$. In general, the reconstruction of an arbitrary density matrix in a $d$-dimensional space requires at least $d^2$ linearly-independent measurements. This requirement can be achieved, for example, by projecting onto a \emph{Symmetric Informationally Complete} (SIC) set of states, or onto a set of \emph{Mutually Unbiased Bases} (MUBs)~\cite{Wootters1989,Derka1998,Scott2006,Zhu2011,bent:2015}. Both measurement strategies ensure complete and non-redundant information acquisition. However, while SIC sets are conjectured (and in some cases constructed) for dimensions up to 100, complete sets of MUBs are known to exist only when the dimension $d$ is a prime number or a power of a prime. For composite dimensions that are not prime powers, such as $d=6$, the existence of a complete set of MUBs remains an open question~\cite{durt:2010,raynal:2011}, posing additional challenges for optimal quantum tomography. Note that, even in non–prime-power dimensions, quantum states can be reconstructed by using overcomplete sets of projections, \emph{e.g.}, through factorisations of the Hilbert space.
Experimentally, the $d^2$ independent elements $\rho_{m,m'}$ of the density matrix, with $m, m' \in \{-\ell, \ldots, +\ell\}$, can be determined through appropriately chosen projective measurements. The diagonal terms $\rho_{m,m}$ are directly accessible as the probabilities of finding the electron in the corresponding OAM eigenstate, \emph{i.e.}, $\rho_{m,m} = \abs{c_m}^2$, where $c_m$ is the amplitude of $\ket{m}$. The off-diagonal terms $\rho_{m,m'}$ (with $m \neq m'$) contain information about the coherences between different OAM modes and must be reconstructed algebraically using measurements that probe superpositions of OAM states. Such coherences can be accessed by projecting onto a set of carefully-chosen superposition states of the form $\frac{1}{\sqrt{2\ell+1}} \sum_{m=-\ell}^{+\ell} e^{i\chi_m} \ket{m}$, where the relative phases $\chi_m$ are varied systematically. These measurements are conceptually analogous to determining \emph{higher-dimensional Stokes parameters} and enable the full recovery of both the amplitudes and the phases that are necessary for complete quantum state tomography in the OAM subspace.\newline
\begin{figure*}
    \centering
    \includegraphics[width=2.0\columnwidth]{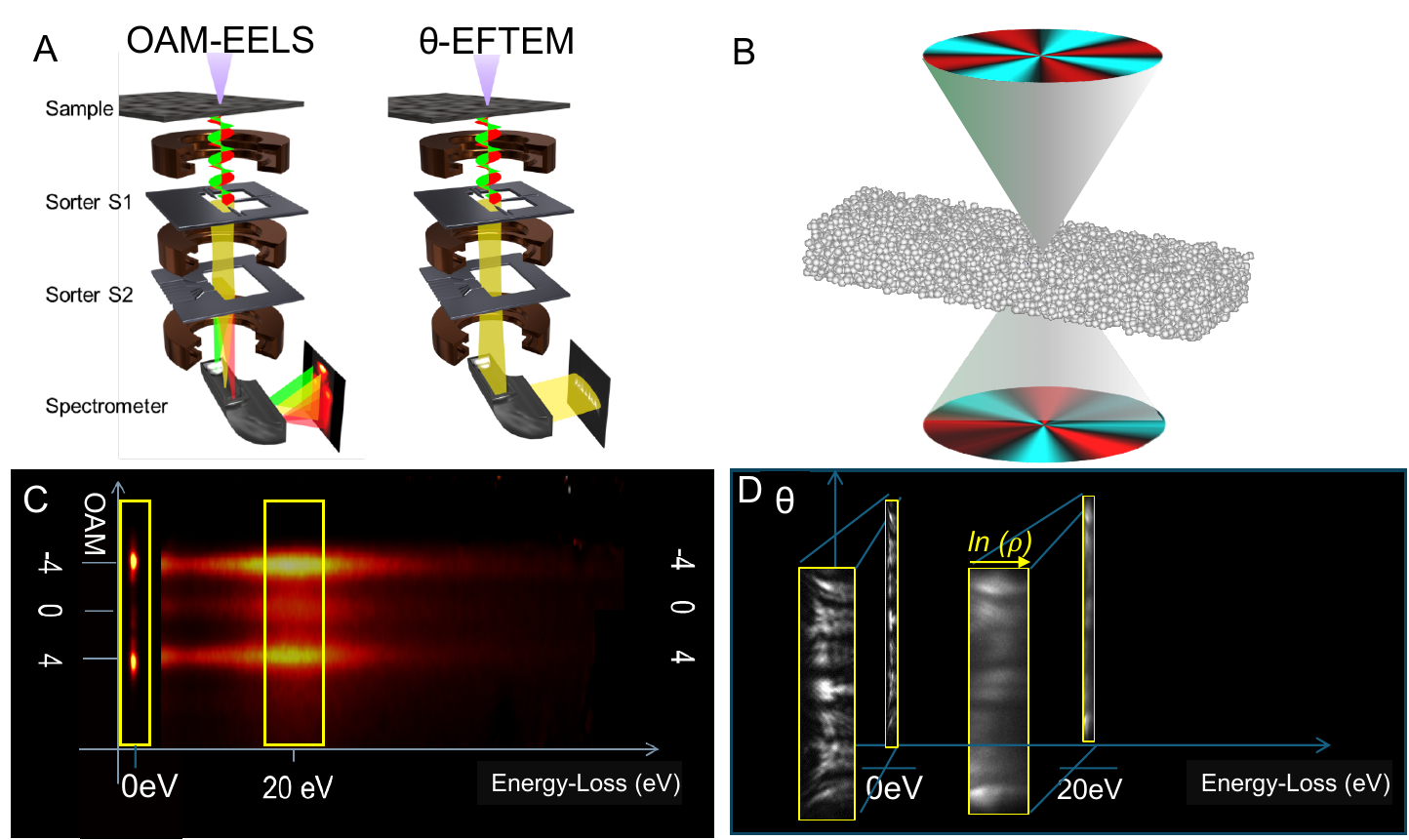}
		\caption{\textbf{(A)} Schematic illustration of the two microscope configurations used for independent measurement of the OAM spectrum and its discrete Fourier transform. By adjusting the excitation of the diffraction lenses, it is possible to transfer either the OAM spectrum or the azimuthal spatial coordinate onto the entrance plane of the spectrometer. \textbf{(B)} Schematic illustration of the experiment: the incident electron beam is a coherent superposition of vortex beams with orbital angular momentum quantum values $m=\pm 4$. 
        \textbf{(C)} Experimental results from OAM-resolved electron energy-loss spectroscopy (OAM-EELS). \textbf{(D)} Experimental energy-filtered transmission electron microscopy (EFTEM) images corresponding to the configuration shown on the right side of \textbf{(A)}. After undergoing a log-polar coordinate transformation, the electron beam is selectively energy-filtered to visualise inelastically-scattered electrons with energy losses near $0$ and $20\,\mathrm{eV}$.} 
		\label{fig:fig1}
\end{figure*}

\noindent{\textbf{OAM state projection:}} 
Several methods can be used for projecting onto specific OAM quantum states, including arbitrary superpositions. A widely-used technique is phase flattening projective measurement, which involves applying a holographic element that imprints the complex conjugate of the target quantum state, effectively implementing the projection $\ket{\psi}\bra{\psi}$, where the hologram encodes $\braket{\psi}{\mathbf{r}} = \psi(\mathbf{r})^*$. After phase conjugation, the relative OAM contribution is given by the on-axis intensity measured in the far field (using, \emph{e.g.}, an aperture, or, in optics experiments, a single mode fiber) ~\cite{karimi:2012,Saitoh:2013,qassim:2014}.
The OAM sorter was first proposed and demonstrated experimentally in the optical domain~\cite{berkhout:2010}. It has subsequently been adapted for electron optics through various implementations~\cite{Grillo2017,Tavabi2021sorter}. Its operation relies on the application of two consecutive phase elements, which are placed in Fourier-conjugate planes (see Fig.~\ref{fig:fig1}-A), to perform a conformal mapping of the transverse wavefunction onto a log-polar coordinate system. Within the geometric optics approximation, the sorter transforms the azimuthal coordinate $\theta$ into a linear coordinate and maps the OAM quantum number $m$ onto conjugate linear momentum.
Unlike the azimuthal coordinate, its conformally-mapped coordinate constitutes a well-defined quantum operator, owing to the topological transformation introduced by the sorter. This topological change involves an unavoidable line of discontinuity that cuts and unwraps the azimuthal coordinate before it is linearised, enabling the construction of legitimate quantum quadratures that are analogous to position and momentum in canonical quantum state tomography.
In the electrostatic realisation of the OAM sorter for electrons, this discontinuity is physically introduced by a \emph{needle-shaped electrode}, which protrudes towards the optical axis. The electrostatic field generated by the needle implements the first phase transformation that is required for the sorter. However, the field discontinuity and associated hard spatial cutoff constitute one of the principal factors that limit the OAM resolution achievable with current designs. Ideally, the OAM sorter should produce a spectrum of discrete peaks, whose full width at half maximum (FWHM) corresponds to exactly one OAM quantum in units of $\hbar$, leading to a certain degree of cross-talk between adjacent OAM modes. Our experimental implementation closely approaches this ideal limit under elastic scattering conditions. More sophisticated designs~\cite{Mirhosseini2013, Rosi2022, Ruffato2018} have been proposed to improve the resolution further and to reduce mode cross-talk, offering pathways for enhanced OAM discrimination in future setups. In an appropriate experimental configuration, OAM measurement can be combined with electron energy-loss spectroscopy (EELS), thereby allowing simultaneous access to the electron’s angular momentum and energy-loss distributions after inelastic scattering~\cite{Bertoni2024}. Alternatively, by selecting a specific energy-loss window using energy-filtered transmission electron microscopy (EFTEM), it is possible to spatially map the inelastically-scattered intensity either in real space (\emph{e.g.}, azimuthal coordinate) or in diffraction space (OAM coordinate). Switching between these two representations requires adjusting the excitation of the diffraction lens to conjugate the energy filter to either of the two planes associated with the OAM sorter. The measurement schemes for OAM and EELS mapping are illustrated in Fig.~\ref{fig:fig1}-A.

In our experiment, the specimen consisted of an amorphous carbon film, and our focus was on investigating the generation and scattering of volume plasmons. As a structured probe, we employed a petal beam, which is a coherent superposition of electron vortex beams with equal and opposite OAM, \emph{i.e.}, states with $|\ell| = 4$. This beam exhibits a characteristic intensity and phase structure with eight-fold rotational symmetry in the transverse plane. The beam was generated using a simple phase hologram (with eight regions possessing phase jumps) placed in the condenser aperture of the microscope, as shown in Fig.~\ref{fig:fig1}-B (see Methods for fabrication and alignment details). The non-trivial structure of the density matrix in the OAM basis following inelastic scattering arises from the complex interaction between this structured electron probe and the collective electronic excitations in the specimen. In this context, plasmon scattering acts as a partial decoherence mechanism and has previously been investigated in simplified scenarios, such as two-slit interference experiments~\cite{verbeeck2005plasmon,harscher1997interference,potapov2007inelastic}, making it a compelling setup for exploring more decoherence and symmetry-breaking questions in high-dimensional OAM states.
Our goal is to examine the mechanism of electron decoherence in the angular/OAM basis. Since energy loss in inelastic scattering varies continuously, our analysis is performed over a finite energy window. The effective density matrix $\rho_{\ell\ell'}$ describing the quantum state in the OAM subspace must be partially traced over both energy and radial degrees of freedom, which we denote generically $r, r'$ and $\Delta E, \Delta E'$, respectively. Formally, this relation can be expressed in the form:
\begin{equation}\label{eq:densitymatrix}
\rho_{\ell\ell'} = \sum_{r, r'} \sum_{\Delta E, \Delta E'} \rho(\ell, \ell'; r, r'; \Delta E, \Delta E'),
\end{equation}
 implying that apparent loss of coherence in the reconstructed density matrix can originate not from the intrinsic inelastic scattering process, but from tracing (integrating) over unresolved degrees of freedom -- here, radial modes and energy. This is a necessary simplification in our analysis, which introduces averaging effects that mask \emph{finer} coherence properties.\newline 

\noindent{\textbf{Experimental results:}}
We selected an energy window of $10\,\mathrm{eV}$ for our measurements. Figure~\ref{fig:fig1}-C shows an OAM-resolved EELS spectrum, in which each vertical slice corresponds to OAM decomposition at a given energy loss. By integrating the doubly-dispersed spectrum along the energy axis in the selected $10\,\mathrm{eV}$ window, a projected OAM distribution is obtained.

Figure~\ref{fig:fig1}-D shows a corresponding EFTEM image of the conformally-transformed electron wave after passing through the OAM sorter and undergoing energy filtering. In this image, the azimuthal coordinate is mapped along the vertical axis, while the radial coordinate lies along the horizontal axis. The radial direction has been traced out, yielding an azimuthal intensity profile that reflects the symmetry of the post-scattering wavefront. If the amorphous carbon sample is considered to be an effective continuous medium, then the scattering geometry is expected to be rotationally-invariant, with any symmetry breaking arising exclusively from the structure of the probe. In our study, OAM becomes a natural and highly informative quantity, which provides a direct probe of the evolution of angular momentum and coherence during inelastic interactions with the sample. We focus our analysis on the joint OAM-energy loss spectrum shown in Fig.~\ref{fig:fig1}-C. On the left side, in the highlighted rectangular region, the zero-loss OAM spectrum is extracted. It shows the expected distribution of an ideal $|m| = 4$ petal beam, with dominant peaks at $m = \pm 4$ and a minimal distribution in other modes. At higher energy losses, the spectrum broadens, and intermediate $m$ values gain in intensity, indicating a redistribution of angular momentum resulting from the inelastic interaction. In order to examine this behaviour, the OAM spectrum slice at each energy loss value was renormalized to its maximum intensity (see Supplementary Figure~S1 for more details). The results confirm that, as the energy loss increases, linear momentum transfer associated with plasmon excitation leads to a degradation of OAM purity. This behaviour is consistent with the expected dispersion relation of plasmons, with higher energy losses corresponding to larger momentum transfer and hence greater perturbation of the probe’s angular momentum state. The use of a structured electron probe enables more profound insight into how symmetry is broken during inelastic scattering. For an OAM eigenstate, transitions between modes do not carry azimuthal information, and the phase is undefined. However, the rotational symmetry is well-defined for a petal beam formed as a superposition of oppositely-charged vortex states, \emph{i.e.}, $\ket{+4}+\ket{-4}$. If the petal beam changes into a new coherent superposition of different OAM states, such as $\ket{5}+\ket{-3}$, then the azimuthal symmetry is broken, but the resulting state possesses alternative information. For instance, the intensity pattern may undergo a rotation in the transverse plane, producing states of the form $\ket{5} + e^{i\chi}\ket{-3}$, where $\chi$ is the relative Gouy phase change and reflects broken symmetry. Alternatively, the superposition may decohere entirely, leading to an incoherent (or partially incoherent) mixture of individual vortex states. Our measurements provide preliminary evidence of such symmetry reduction and OAM exchange, laying the groundwork for future investigations into coherence-preserving and coherence-breaking mechanisms in structured electron beams.
\begin{figure*}
    \centering
    \includegraphics[width=0.9\textwidth]{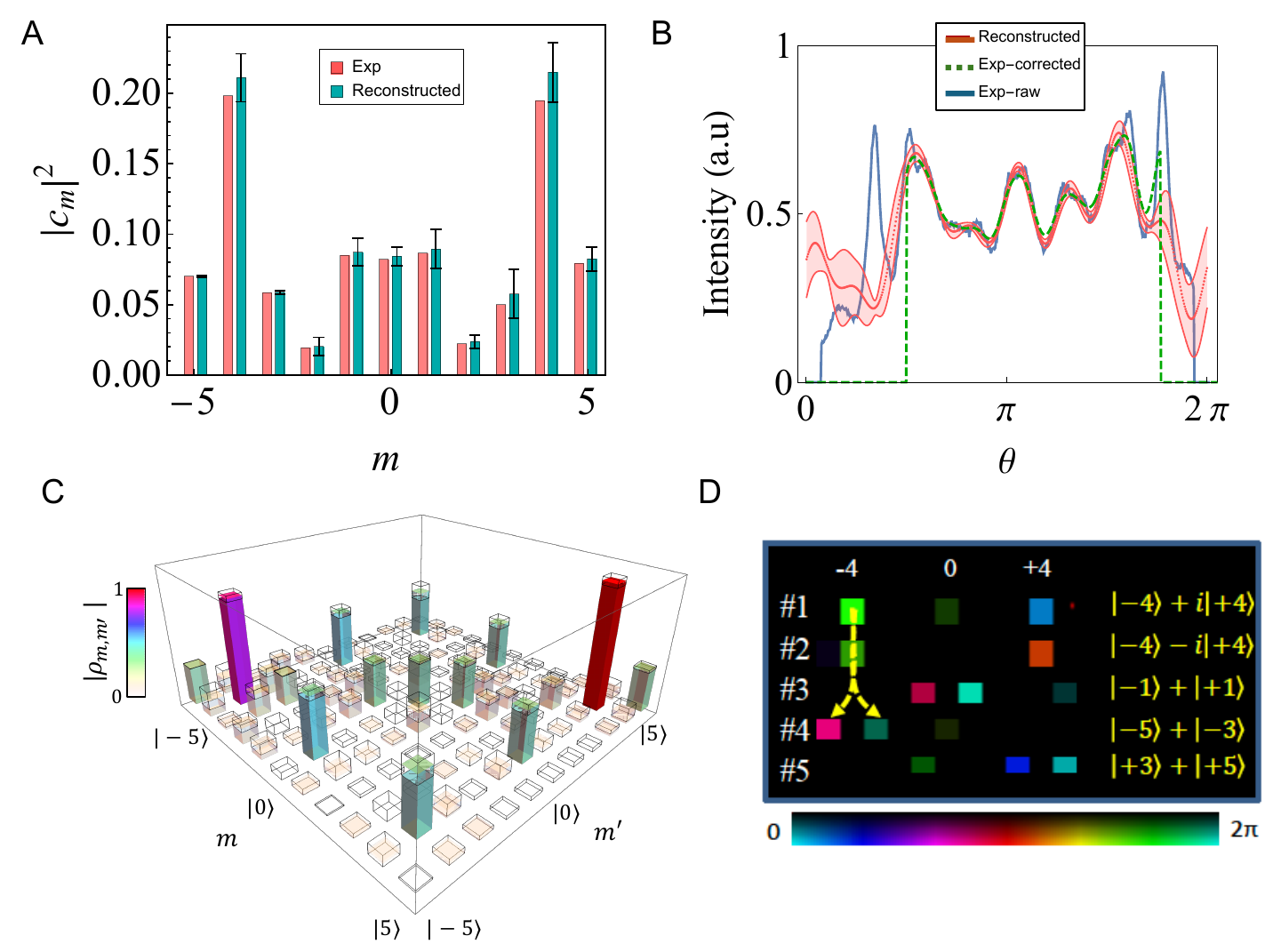}
		\caption{\textbf{(A)} Experimental OAM spectrum after inelastic scattering, obtained by integrating the energy-resolved data from Fig.~\ref{fig:fig1}-C over a $10\,\mathrm{eV}$ window. The measured distribution is shown alongside the best-fitting result from a constrained maximum-likelihood estimation (MLE) algorithm used to reconstruct the OAM density matrix. \textbf{(B)} Azimuthal intensity profile corresponding to the same energy window, obtained by averaging the EFTEM image over the radial coordinate (\emph{c.f.} Fig.~\ref{fig:fig1}-D). The experimental profile (red) is shown together with the angular projection derived from the fitted density matrix. \textbf{(C)} Reconstructed  modulus of the OAM density matrix $|\hat{\rho}|$, visualized in the truncated basis with $|\ell| \leq 5$. The matrix represents the best-fitting solution subject to constraints of positivity, symmetry, and elastic-state proximity. \textbf{(D)} Graphical representation of the five dominant eigenstates of $\hat{\rho}$, ranked by eigenvalue. For each state, a qualitative decomposition into OAM basis states is illustrated. The full complex coefficients are provided in the Supplementary Information. Errors correspond to 3-times the standard deviation over 10 reconstructed density matrices.} 
		\label{fig:fig2}
\end{figure*}

Combining experimental tomographic reconstructions with model-based analysis can provide deeper insight into the mechanisms underlying angular momentum redistribution and coherence loss in inelastic electron scattering. Below, we analyse the tomographically-reconstructed quantum state in the principal energy window, focusing on both the OAM and the angular representation. Figure~\ref{fig:fig1}-D shows the experimentally-measured angular intensity profiles. The profiles have a dominant central modulation and two prominent side lobes, which we attribute, in part, to the geometric cutoff introduced by the needle electrode used in the OAM sorter (S1 in Fig.~\ref{fig:fig1} A), which partially obstructs the incoming beam. When comparing the angular profiles for the elastic and inelastic scattering conditions, we find that they are broadly similar in their structure. They both exhibit a primary oscillatory pattern with seven visible fringes. Whereas the petal beam associated with the initial $|m| = 4$ state nominally has eight angular lobes, one of them is occluded by the shadow of the sorter’s electrode. A four-fold amplitude modulation is also superimposed, arising from imperfections or residual structure in the phase hologram that was used to generate the beam. However, a key difference emerges in the fine structure of the profiles. In the elastic case, higher-order angular modulations are visible, indicating the presence of long-range coherence and fine phase structure. These higher-order features are suppressed in the inelastic case, suggesting that inelastic scattering introduces greater decoherence, particularly at higher spatial frequencies. Importantly, these fine features lie beyond the resolution accessible in our current analysis due to limitations imposed by the truncated OAM basis. For density matrix reconstruction, we restricted our analysis to a finite-dimensional vector space comprising $|\ell| \leq 5$, $d=2\ell+1=11$, resulting in an $11 \times 11$ matrix representation. Within this constrained basis, the density matrix $\hat{\rho}$ includes 120 independent real parameters (accounting for Hermiticity and normalisation, \emph{i.e.}, $\mathrm{Tr}(\hat{\rho}) = 1$). The number of independent experimental observables remains considerably smaller: we extract 11~coefficients from the OAM-resolved EELS data and approximately 21 from the angular intensity profile. The inverse problem is therefore under-determined. We introduce additional \emph{a~priori} information based on physical and experimental considerations to mitigate this indeterminacy. First, we assume that the quantum state should be approximately symmetrical with respect to positive and negative values of $m$, reflecting the initial structure of the probe and the approximate symmetry of the scattering geometry. Minor asymmetries may arise from misalignments or imperfections in the optical system, but we impose symmetry as a soft constraint. More importantly, we incorporate the knowledge that the post-scattering state should remain close, in trace distance, to the nearly pure state observed in the elastic scattering case.

In order to solve the inverse problem under these constraints, we developed a modified maximum-likelihood estimation (MLE) algorithm for quantum state reconstruction, which incorporates Lagrange multipliers to enforce soft constraints \emph{via} a weighted cost function~\cite{james:2001}. These constraints include positivity, trace normalisation, approximate $m$-symmetry, and proximity to the known elastic-state density matrix. Details of the reconstruction algorithm and regularization strategy are provided in the Methods section.

The reconstructed modulus of the density matrix $\hat{\rho}$ is shown in Fig.~\ref{fig:fig2}-C, together with corresponding fits to the experimental OAM and angular intensity profiles (Fig.~\ref{fig:fig2}-A and \ref{fig:fig2}-B). A useful scalar quantity for characterising the degree of quantum coherence in a system is purity ${\cal P} = \mathrm{Tr}(\hat{\rho}^2)$. This quantity ranges from ${\cal P} = 1$ for a completely pure state to ${\cal P} = 1/d^2$ (or effectively 0 for large $d$) for a fully mixed, incoherent state.  In our case, we observed a substantial decrease in purity following inelastic scattering. We measured a purity of ${\cal P} = 0.54$ for the elastic probe state, which was reduced to ${\cal P}= 0.21$ after inelastic interactions. It is worth noting that even the initial probe is not strictly pure, as averaging over unresolved radial degrees of freedom leads to partial decoherence in the OAM basis. The purity would approach unity if the radial and azimuthal modes were measured simultaneously. Nevertheless, the reduction in ${\cal P}$ after scattering indicates a significant loss of coherence. Despite this loss, a non-negligible degree of coherence persists after inelastic scattering. This observation is consistent with the persistence of low frequency fringes in the experimental angular intensity profiles (\emph{c.f.} Fig.~\ref{fig:fig1}-D), which confirm that coherence is not lost entirely. In passing, we also note that the conformal transformation performed by the OAM sorter -- despite being rooted in coherent wave optics -- appears to remain valid for partially coherent beams, further validating its use in quantum state tomography (QST).
In order to further analyse the physical content of the reconstructed density matrix, we diagonalised $\hat{\rho}$. In this representation, the quantum state is decomposed into a weighted sum of orthogonal pure states (eigenstates), each of which is associated with a corresponding eigenvalue that represents its occupation probability. This representation is not only mathematically convenient, but also offers physical insight. As shown in the Supplementary Information, direct correspondence exists between these eigenstates and the final states of the system following a perturbative inelastic interaction. Specifically, the eigenstates of $\hat{\rho}$ can be interpreted as possible post-scattering states, as defined by Fermi’s golden rule, while their eigenvalues quantify the transition probability into each state. This interpretation provides a powerful new way to analyse electron energy-loss spectroscopy (EELS) using QST: rather than observing only the integrated signal or a specific final state, one can retrieve the full ensemble of post-interaction states \emph{via} diagonalisation of the measured density matrix. Furthermore, when the interaction potential is known or constrained by conservation laws, this formalism allows the observed transitions in the electron’s state to be related directly to transitions in the sample.
Figure~\ref{fig:fig2}-D shows a graphical representation of the five most significant eigenstates of the reconstructed density matrix. Each row corresponds to a pure eigenstate, while each dot represents its decomposition in the OAM basis. The marker brightness reflects the amplitude $|c_m|^2$, while color encodes the complex phase of the OAM component in the state $\psi(\theta) = \sum_{m} c_m e^{im\theta}$. A complete complex decomposition of all eigenstates is provided in the Supplementary Information. Several important observations emerge from this analysis. First, the dominant eigenstate closely matches the input probe state, which is consistent with the reconstruction constraint that the post-scattering state remains close to the elastic reference. Nevertheless, its clear emergence as the leading eigenstate validates the reconstruction and confirms that this mode survives the interaction with relatively high coherence. The second eigenstate has a similar OAM composition but a different azimuthal phase structure, suggesting that it arises from a coherent rotation of the probe in momentum space, effectively corresponding to a shifted angular origin. The nature of this transformation will be explored further below. Intriguingly, the third, fourth, and fifth eigenstates reveal transitions from the initial probe through single-OAM-step changes, \emph{i.e.}, $\Delta m = \pm 1$, from each dominant component of the petal beam. This observation aligns with the expected dipole selection rules and suggests that the inelastic electron-plasmon interaction is governed predominantly by dipolar coupling. This result was not imposed as a prior assumption in the reconstruction and arises as an outcome of data-driven density matrix analysis. In the Supplementary Information, we analyse higher-order eigenstates of $\hat{\rho}$ and identify weaker signatures corresponding to $\Delta m = \pm 2$ quadrupole transitions, which can be automatically separated through eigenmode decomposition, offering a pathway for the systematic identification of multipolar contributions in future high-resolution experiments. Finally, we observe a slight asymmetry between the positive and negative $\ell$ components, even though our reconstruction enforced a weak symmetry constraint. This small deviation may be attributed to incomplete convergence of the optimisation algorithm, or to minor experimental imperfections such as misalignment of the OAM sorter.
\begin{figure*}
    \centering
    \includegraphics[width=1.8\columnwidth]{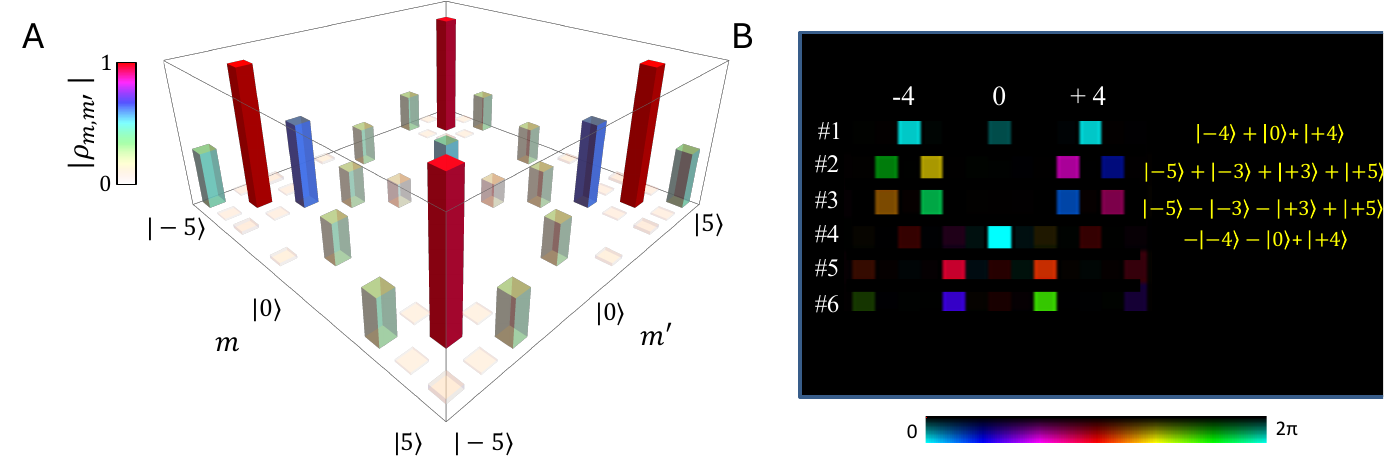}
		\caption{\textbf{(A)} Simulated density matrix obtained using the Monte Carlo model of momentum exchange, with inelastic scattering treated as a random lateral shift in momentum space. The OAM-resolved representation captures decoherence and the redistribution of angular momentum. \textbf{(B)} Eigenstate decomposition of the simulated density matrix. The model produces degenerate eigenstates with identical OAM spectral content but differing azimuthal phases. The states are rotated versions of one another, preserving rotational invariance in spite of symmetry-breaking features introduced by the plasmon-mediated interaction.} 
		\label{fig:fig3}
\end{figure*}

In order to better understand plasmon excitation dynamics, we developed a simulation framework based on a Monte Carlo approach. In this model, the inelastic interaction is assumed to involve the exchange of plane waves between the electron and the plasmon field, without incorporating delocalisation \cite{egerton2007electron} or shape factors. This approach enables a compact analytical treatment that focuses on the formal structure of the resulting density matrix.
Specifically, the OAM-resolved density matrix as a function of energy loss $\Delta {\cal E}$ is given by the expression:
\begin{eqnarray}
\rho_{m,m’}(\Delta {\cal E})&=&\int_{0}^{|\mathbf{k}|< k_\mathrm{p,max}(\Delta {\cal E})} d|\mathbf{k}_\mathrm{p}|\,\,\\\nonumber
&\times& G(\mathbf{k}_\mathrm{p})\, \delta(\mathbf{q} - \mathbf{k}_\mathrm{p})\bra{m} e^{-i \mathbf{q} \cdot \mathbf{r}_e} \ket{\psi_\mathrm{ei}} 
\bra{\psi_\mathrm{ei}} e^{i \mathbf{q} \cdot \mathbf{r}_e} \ket{m'}
,
\end{eqnarray}
where $\ket{\psi_\mathrm{ei}}$ is the initial probe state, and we explicitly use the fact that the density matrix is diagonal in the transferred momentum components $\mathbf{q}, \mathbf{q}'$ (see the Supplementary Material for a derivation). We model the plasmon as a perfect in-plane plane wave with momentum $\mathbf{k}_\mathrm{p}$, such that $\mathbf{q} = \mathbf{k}_\mathrm{p}$ corresponds to the momentum exchanged with the electron during inelastic scattering. 
This formulation ignores both the delocalisation of the interaction~\cite{egerton2007electron} and the internal spatial structure of the plasmon, treating it instead as a dispersion-limited interaction with a momentum cutoff $k_\mathrm{p,max}(\Delta {\cal E})$ determined by the energy loss $\Delta {\cal E}$. We also average over all radial momentum components up to this cutoff. Despite its simplicity, this model provides insight into the structure of the density matrix and its eigenstates. The first eigenstate corresponds to the unperturbed probe state, as in the experimental data. The second and third eigenstates are nearly degenerate and exhibit similar OAM distributions, but differ in their azimuthal phase. These two states are connected \emph{via} a rotation in angular momentum space and emerge naturally from the dipole selection rules applied to the initial structured state. The rotational invariance of the interaction permits arbitrary superpositions of these degenerate states. In other words, any linear combination of the second and third eigenstates -- including states with arbitrary rotational orientation -- represents a valid post-interaction state, reflecting the absence of a preferential direction for angular momentum transfer in the system. The symmetry of the interaction thus preserves angular isotropy, while allowing transitions that change OAM. In this picture, the model resolves a previously noted conceptual tension between maintaining rotational symmetry and observing OAM exchange in inelastic scattering~\cite{larocque2016mapping}. Because the system supports degenerate final states that are related by rotation, there is no contradiction between symmetry and angular momentum transfer.
Furthermore, higher-order eigenstates of the simulated density matrix, such as eigenstates 5 and 6, exhibit $\Delta m = \pm 2$ transitions, which correspond to quadrupole coupling and become especially visible in the simulation due to the controlled conditions and full knowledge of the matrix elements for each $m$ value. This analysis demonstrates that simplified Monte Carlo simulations -- when integrated with quantum state tomography -- can be a powerful tool for interpreting and predicting multipolar dynamics in structured electron-matter interactions. Our experimental findings reveal an approximate degeneracy between the second and third eigenvalues of the reconstructed density matrix. This small splitting may lie within the uncertainty limits of the reconstruction process, particularly when considering possible minor misalignments of the OAM sorter. If the symmetry breaking is real, then it may be attributed to the finite spatial extent of the plasmonic excitation, which deviates from an ideal plane-wave model. In such a case (considering the intrinsic delocalisation of the plasmon-electron interaction), coupling with a tightly-focused probe could induce slight asymmetry in the emitted field -- manifesting as a preferred direction for momentum exchange or angular deflection. Strikingly, both the simulations and the experiment indicate that dipolar transitions -- corresponding to $\Delta m = \pm 1$ -- occur coherently and simultaneously. From the perspective of angular momentum conservation, this observation suggests that the plasmon is emitted in a near-plane-wave configuration. Furthermore, the use of an OAM sorter allows post-selection of specific angular momentum values. In principle, one could isolate transitions corresponding to $\Delta m= +1$, while suppressing those with $\Delta m = -1$, implying the generation of a plasmon with non-zero topological charge. Such a process would correspond to a highly exotic vortex-like excitation in the plasmonic field -- an otherwise difficult state to prepare directly.
These observations are related to a fundamental principle of quantum measurement: when the total state of a system is entangled, the \qo{collapse} of one subsystem enforces a corresponding \qo{collapse} of the other. In the context of electron-plasmon interactions, post-selection on the electron’s final OAM state effectively prepares the plasmon in a correlated vortex state. Whereas volume plasmons are typically short-lived, the case of surface plasmon polaritons (SPPs) presents a more promising avenue. As a result of their longer coherence lengths and surface-bound nature, the post-selection of an electron’s OAM could be used to prepare vortex modes in an SPP field, thereby offering a new approach to the controlled generation of structured collective excitations.
\section{Conclusions}
In conclusion, this work demonstrates a powerful shift in the use of energy-loss spectroscopy in the transmission electron microscope by introducing quantum state tomography as an intrinsic analysis tool for inelastic electron scattering. Full EELS quantum-state tomography is achieved in a discrete orbital-angular-momentum basis. Rather than relying on a pre-defined set of projective measurements, complete reconstruction of the post-interaction density matrix enables an eigenstate decomposition that provides a basis-independent description of the final electron state. This decomposition provides direct access to the full structure of the interaction. Each eigenstate corresponds to a physically admissible transition channel, while the associated eigenvalue quantifies its statistical weight. The formalism mirrors the logic of Fermi’s golden rule, but without requiring \emph{a~priori} specification of the relevant final states. As a case study, the coherence loss of an electron exciting a volume plasmon in amorphous carbon is analyzed. The tomographic reconstruction automatically recovers the expected dipole (and more generally multipole) selection rules, while simultaneously revealing the spontaneous breaking of rotational symmetry encoded in the mixed final state.
Beyond identifying allowed transitions, the density matrix formalism makes the mechanisms of coherence loss experimentally accessible. The redistribution of weight among the eigenstates provides a quantitative and operational measure of decoherence induced by the inelastic interaction, independent of the measurement basis. Quantum state tomography in the OAM space is identified as a new observable in EELS, offering a structurally complete description of excitation processes in matter that unifies selection rules, symmetry considerations, and coherence properties within a single experimental framework.


\vspace{0.5cm}
\noindent \textbf{Acknowledgments.}
A.D and E.K. acknowledge the support of the Canada Research Chairs and the Alliance Consortia Quantum Grant (QuEnSI). \\

\noindent \textbf{Author contributions}
A.H.T., A.D., and P.R. contributed equally to this work. E.K. conceived the concept of quantum tomography and the numerical methods employed. V.G. conceived the application, designed the experiment, and developed the diagonalisation interpretation. A.H.T. performed all experiments, including OAM alignment. A.D.E. applied the numerical methods. P.R. designed and fabricated all holographic masks, optimised the MEMS devices, and contributed to the experimental work. G.B. contributed to the interpretation of the EELS measurements. E.R. performed optimisation and simulations of the sorter. L.B. and A.R. fabricated the MEMS devices for the sorter. S.F. assisted with the sorter setup. G.C.G. contributed to the refinement of the MEMS structures. P.T. contributed to the development of the sorter electron optics. R.E.D-B. contributed to the data analysis and interpretation. E.K. and V.G. supervised the project. All of the authors discussed the results and contributed to writing the manuscript. \\

\noindent \textbf{Disclosures} The authors declare no conflicts of interest.\\

\noindent \textbf{Data availability} Data underlying the results presented in this paper may be obtained from the authors upon reasonable request.


\bibliographystyle{unsrt}
\bibliography{EOAMST}

@preamble{"\providecommand{\noopsort}[1]{}"}

@book{Cohen-Tannoudji1991,
author = {Cohen-Tannoudji, Claude and Diu, Bernard and Lalo{\"{e}}, Franck},
isbn = {978-0471164333},
pages = {914},
publisher = {Wiley},
title = {{Quantum Mechanics}},
year = {1991}
}

@article{Gerchberg1972,
author = {Gerchberg, R.W. and Saxton, W.O.},
issn = {1063-7818},
journal = {Optik},
number = {2},
pages = {237--246},
title = {{A Practical Algorithm for the Determination of Phase from Image and Diffraction Plane Pictures}},
volume = {35},
year = {1972}
}

@article{Fienup1982,
author = {Fienup, J. R.},
doi = {10.1364/AO.21.002758},
issn = {0003-6935},
journal = {Applied Optics},
month = {aug},
number = {15},
pages = {2758},
title = {{Phase retrieval algorithms: a comparison}},
url = {https://opg.optica.org/abstract.cfm?URI=ao-21-15-2758},
volume = {21},
year = {1982}
}

@article{lundeen2011direct,
  title={Direct measurement of the quantum wavefunction},
  author={Lundeen, Jeff S and Sutherland, Brandon and Patel, Aabid and Stewart, Corey and Bamber, Charles},
  journal={Nature},
  volume={474},
  number={7350},
  pages={188--191},
  year={2011},
  publisher={Nature Publishing Group UK London}
}

@article{liberman2016quantum,
  title={Quantum enhanced phase retrieval},
  author={Liberman, Liat and Israel, Yonatan and Poem, Eilon and Silberberg, Yaron},
  journal={Optica},
  volume={3},
  number={2},
  pages={193--199},
  year={2016},
  publisher={Optical Society of America}
}

@article{PhysRevLett.105.150401,
  title = {Quantum State Tomography via Compressed Sensing},
  author = {Gross, David and Liu, Yi-Kai and Flammia, Steven T. and Becker, Stephen and Eisert, Jens},
  journal = {Phys. Rev. Lett.},
  volume = {105},
  issue = {15},
  pages = {150401},
  numpages = {4},
  year = {2010},
  month = {Oct},
  publisher = {American Physical Society},
  doi = {10.1103/PhysRevLett.105.150401},
  url = {https://link.aps.org/doi/10.1103/PhysRevLett.105.150401}
}

@article{compressivesensing,
  title = {Adaptive Compressive Tomography with No a priori Information},
  author = {Ahn, D. and Teo, Y. S. and Jeong, H. and Bouchard, F. and Hufnagel, F. and Karimi, E. and Koutn\'y, D. and \ifmmode \check{R}\else \v{R}\fi{}eh\'a\ifmmode \check{c}\else \v{c}\fi{}ek, J. and Hradil, Z. and Leuchs, G. and S\'anchez-Soto, L. L.},
  journal = {Phys. Rev. Lett.},
  volume = {122},
  issue = {10},
  pages = {100404},
  numpages = {5},
  year = {2019},
  month = {Mar},
  publisher = {American Physical Society},
  doi = {10.1103/PhysRevLett.122.100404},
  url = {https://link.aps.org/doi/10.1103/PhysRevLett.122.100404}
}

@article{dehghan2024biphoton,
  title={Biphoton state reconstruction via phase retrieval methods},
  author={Dehghan, Nazanin and D’Errico, Alessio and Di Colandrea, Francesco and Karimi, Ebrahim},
  journal={Optica},
  volume={11},
  number={8},
  pages={1115--1123},
  year={2024},
  publisher={Optica Publishing Group}
}

@article{schattschneider1999density,
  title={Density matrix of inelastically scattered fast electrons},
  author={Schattschneider, P. and Nelhiebel, M. and Jouffrey, B.},
  journal={Physical Review B},
  volume={59},
  number={16},
  pages={10959--10969},
  year={1999},
  publisher={APS}
}

@article{schattschneider2000physical,
  title={The physical significance of the mixed dynamic form factor},
  author={Schattschneider, P. and Nelhiebel, M. and Souchay, H. and Jouffrey, B.},
  journal={Micron},
  volume={31},
  number={4},
  pages={333--345},
  year={2000},
  publisher={Elsevier}
}

@article{rodenburg1992ptychography,
  title={Ptychography and related diffractive imaging methods},
  author={Rodenburg, A. J.},
  journal={Philosophical Transactions of the Royal Society of London. Series A: Physical and Engineering Sciences},
  volume={339},
  number={1655},
  pages={521--553},
  year={1992},
  publisher={The Royal Society}
}

@article{Jiang2018,
author = {Jiang, Yi and Chen, Zhen and Han, Yimo and Deb, Pratiti and Gao, Hui and Xie, Saien and Purohit, Prafull and Tate, Mark W. and Park, Jiwoong and Gruner, Sol M. and Elser, Veit and Muller, David A.},
doi = {10.1038/s41586-018-0298-5},
issn = {0028-0836},
journal = {Nature},
month = {jul},
number = {7714},
pages = {343--349},
title = {{Electron ptychography of 2D materials to deep sub-{\aa}ngstr{\"{o}}m resolution}},
url = {http://www.nature.com/articles/s41586-018-0298-5 https://www.nature.com/articles/s41586-018-0298-5},
volume = {559},
year = {2018}
}

@article{Yang2017,
title = {Electron ptychographic phase imaging of light elements in crystalline materials using Wigner distribution deconvolution},
journal = {Ultramicroscopy},
volume = {180},
pages = {173-179},
year = {2017},
note = {Ondrej Krivanek: A research life in EELS and aberration corrected STEM},
issn = {0304-3991},
doi = {https://doi.org/10.1016/j.ultramic.2017.02.006},
url = {https://www.sciencedirect.com/science/article/pii/S0304399117300773},
author = {Hao Yang and Ian MacLaren and Lewys Jones and Gerardo T. Martinez and Martin Simson and Martin Huth and Henning Ryll and Heike Soltau and Ryusuke Sagawa and Yukihito Kondo and Colin Ophus and Peter Ercius and Lei Jin and András Kovács and Peter D. Nellist},
}

@article{Roder2014,
author = {R{\"{o}}der, Falk and Lubk, Axel},
doi = {10.1016/j.ultramic.2014.07.007},
issn = {03043991},
journal = {Ultramicroscopy},
month = {nov},
pages = {103--116},
title = {{Transfer and reconstruction of the density matrix in off-axis electron holography}},
url = {https://linkinghub.elsevier.com/retrieve/pii/S0304399114001466},
volume = {146},
year = {2014}
}

@article{priebe2017attosecond,
  title={Attosecond electron pulse trains and quantum state reconstruction in ultrafast transmission electron microscopy},
  author={Priebe, K. E. and Rathje, C. and Yalunin, S. V. and Hohage, T. and Feist, A. and Schafer, S. and Ropers, C.},
  journal={Nature Photonics},
  volume={11},
  pages={793--797},
  year={2017},
  publisher={Nature Publishing Group}
}

@article{gaida2024attosecond,
  title={Attosecond electron microscopy by free-electron homodyne detection},
  author={Gaida, J. H. and Lourenço-Martins, H. and Sivis, M. and Yalunin, S. V. and Feist, A. and Ropers, C.},
  journal={Nature Photonics},
  volume={18},
  pages={509--515},
  year={2024},
  publisher={Nature Publishing Group}
}

@article{bucher2023free,
  title={Free-electron Ramsey-type interferometry for enhanced amplitude and phase imaging of nearfields},
  author={Bucher, T. and Liao, G. and Ropers, C. and Feist, A.},
  journal={Science Advances},
  volume={9},
  number={eadi5729},
  year={2023},
  publisher={American Association for the Advancement of Science}
}

@article{iwashimizu2021electron,
  title={Electron orbital mapping of SrTiO3 using electron energy-loss spectroscopy},
  author={Iwashimizu, C. and Haruta, M. and Kurata, H.},
  journal={Applied Physics Letters},
  volume={119},
  number={23},
  pages={232902},
  year={2021},
  publisher={AIP Publishing}
}

@article{haas2022advances,
  title={Advances in electron microscopy techniques},
  author={Haas, B. and Koch, C. T.},
  journal={Microscopy and Microanalysis},
  volume={28},
  number={S1},
  pages={1--2},
  year={2022},
  publisher={Cambridge University Press}
}

@article{Tavabi2021sorter,
author = {Tavabi, Amir H. and Rosi, Paolo and Rotunno, Enzo and Roncaglia, Alberto and Belsito, Luca and Frabboni, Stefano and Pozzi, Giulio and Gazzadi, Gian Carlo and Lu, Peng-Han and Nijland, Robert and Ghosh, Moumita and Tiemeijer, Peter and Karimi, Ebrahim and Dunin-Borkowski, Rafal E. and Grillo, Vincenzo},
doi = {10.1103/PhysRevLett.126.094802},
eprint = {1910.03706},
issn = {0031-9007},
journal = {Physical Review Letters},
month = {mar},
number = {9},
pages = {094802},
publisher = {American Physical Society},
title = {{Experimental Demonstration of an Electrostatic Orbital Angular Momentum Sorter for Electron Beams}},
url = {https://doi.org/10.1103/PhysRevLett.126.094802 https://link.aps.org/doi/10.1103/PhysRevLett.126.094802 http://arxiv.org/abs/1910.03706},
volume = {126},
year = {2021}
}

@article{Grillo2017,
author = {Grillo, Vincenzo and Tavabi, Amir H. and Venturi, Federico and Larocque, Hugo and Balboni, Roberto and Gazzadi, Gian Carlo and Frabboni, Stefano and Lu, Peng-Han and Mafakheri, Erfan and Bouchard, Fr{\'{e}}d{\'{e}}ric and Dunin-Borkowski, Rafal E. and Boyd, Robert W. and Lavery, Martin P. J. and Padgett, Miles J. and Karimi, Ebrahim},
doi = {10.1038/ncomms15536},
issn = {2041-1723},
journal = {Nature Communications},
month = {aug},
number = {1},
pages = {15536},
title = {{Measuring the orbital angular momentum spectrum of an electron beam}},
url = {http://www.nature.com/articles/ncomms15536},
volume = {8},
year = {2017}
}

@article{Mirhosseini2013,
author = {Mirhosseini, Mohammad and Malik, Mehul and Shi, Zhimin and Boyd, Robert W.},
doi = {10.1038/ncomms3781},
issn = {2041-1723},
journal = {Nature Communications},
month = {dec},
number = {1},
pages = {2781},
title = {{Efficient separation of the orbital angular momentum eigenstates of light}},
url = {http://www.nature.com/articles/ncomms3781},
volume = {4},
year = {2013}
}

@article{Rosi2022,
author = {Rosi, Paolo and Venturi, Federico and Medici, Giacomo and Menozzi, Claudia and Gazzadi, Gian Carlo and Rotunno, Enzo and Frabboni, Stefano and Balboni, Roberto and Rezaee, Mohammadreza and Tavabi, Amir H. and Dunin-Borkowski, Rafal E. and Karimi, Ebrahim and Grillo, Vincenzo},
doi = {10.1063/5.0067528},
eprint = {2109.11347},
issn = {0021-8979},
journal = {Journal of Applied Physics},
month = {jan},
number = {3},
pages = {031101},
title = {{Theoretical and practical aspects of the design and production of synthetic holograms for transmission electron microscopy}},
url = {http://arxiv.org/abs/2109.11347 https://aip.scitation.org/doi/10.1063/5.0067528},
volume = {131},
year = {2022}
}

@article{Ruffato2018,
author = {Ruffato, Gianluca and Girardi, Marcello and Massari, Michele and Mafakheri, Erfan and Sephton, Bereneice and Capaldo, Pietro and Forbes, Andrew and Romanato, Filippo},
doi = {10.1038/s41598-018-28447-1},
issn = {2045-2322},
journal = {Scientific Reports},
month = {jul},
number = {1},
pages = {10248},
title = {{A compact diffractive sorter for high-resolution demultiplexing of orbital angular momentum beams}},
url = {https://www.nature.com/articles/s41598-018-28447-1},
volume = {8},
year = {2018}
}

@misc{Bertoni2024,
author = {Bertoni, Giovanni and Tavabi, Amir and Rosi, Paolo and Rotunno, Enzo and Belsito, Luca and Roncaglia, Alberto and Frabboni, Stefano and Karimi, Ebrahim and Tiemeijer, Peter and Dunin-Borkowski, Rafal and Grillo, Vincenzo and Gazzadi, Gian Carlo},
doi = {10.21203/rs.3.rs-4606043/v1},
institution = {Research Square},
month = {jul},
title = {{First demonstration of angular-momentum-resolved electron energy-loss spectroscopy}},
url = {https://www.researchsquare.com/article/rs-4606043/v1},
year = {2024}
}

@article{karimi:2012,
  title={Radial coherent and intelligent states<? A3B2 show [pmg: line-break justify=" yes"/]?> of paraxial wave equation},
  author={Karimi, Ebrahim and Santamato, Enrico},
  journal={Optics letters},
  volume={37},
  number={13},
  pages={2484--2486},
  year={2012},
  publisher={Optical Society of America}
}

@article{karimi:2014a,
  title={Radial quantum number of Laguerre-Gauss modes},
  author={Karimi, E and Boyd, RW and De La Hoz, P and De Guise, H and {\v{R}}eh{\'a}{\v{c}}ek, J and Hradil, Z and Aiello, A},
  journal={Physical review A},
  volume={89},
  number={6},
  pages={063813},
  year={2014},
  publisher={APS}
}

@article{karimi:2014b,
  title={Exploring the quantum nature of the radial degree of freedom of a photon via Hong-Ou-Mandel interference},
  author={Karimi, Ebrahim and Giovannini, Daniel and Bolduc, Eliot and Bent, Nicolas and Miatto, Filippo M and Padgett, Miles J and Boyd, Robert W},
  journal={Physical Review A},
  volume={89},
  number={1},
  pages={013829},
  year={2014},
  publisher={APS}
}

@article{plick:2015,
  title={Physical meaning of the radial index of Laguerre-Gauss beams},
  author={Plick, William N and Krenn, Mario},
  journal={Physical Review A},
  volume={92},
  number={6},
  pages={063841},
  year={2015},
  publisher={APS}
}

@article{bliokh:2023,
  title={Roadmap on structured waves},
  author={Bliokh, Konstantin Y and Karimi, Ebrahim and Padgett, Miles J and Alonso, Miguel A and Dennis, Mark R and Dudley, Angela and Forbes, Andrew and Zahedpour, Sina and Hancock, Scott W and Milchberg, Howard M and others},
  journal={Journal of Optics},
  volume={25},
  number={10},
  pages={103001},
  year={2023},
  publisher={IOP Publishing}
}

@article{durt:2010,
  title={On mutually unbiased bases},
  author={Durt, Thomas and Englert, Berthold-Georg and Bengtsson, Ingemar and {\.Z}yczkowski, Karol},
  journal={International journal of quantum information},
  volume={8},
  number={04},
  pages={535--640},
  year={2010},
  publisher={World Scientific}
}

@article{raynal:2011,
  title={Mutually unbiased bases in six dimensions: The four most distant bases},
  author={Raynal, Philippe and L{\"u}, Xin and Englert, Berthold-Georg},
  journal={Physical Review A—Atomic, Molecular, and Optical Physics},
  volume={83},
  number={6},
  pages={062303},
  year={2011},
  publisher={APS}
}

@article{Wootters1989,
  author       = {W. K. Wootters and B. D. Fields},
  title        = {Optimal State-Determination by Mutually Unbiased Measurements},
  journal      = {Annals of Physics},
  volume       = {191},
  number       = {2},
  pages        = {363--381},
  year         = {1989},
  doi          = {10.1016/0003-4916(89)90322-9}
}

@article{Derka1998,
  author       = {R. Derka and V. Bu{\v{z}}ek and A. Ekert},
  title        = {Universal Algorithm for Optimal Estimation of Quantum States from Finite Ensembles via Realizable Generalized Measurement},
  journal      = {Physical Review Letters},
  volume       = {80},
  number       = {8},
  pages        = {1571--1575},
  year         = {1998},
  doi          = {10.1103/PhysRevLett.80.1571}
}

@article{Scott2006,
  author       = {A. J. Scott},
  title        = {Tight Informationally Complete Quantum Measurements},
  journal      = {Journal of Physics A: Mathematical and General},
  volume       = {39},
  number       = {43},
  pages        = {13507--13530},
  year         = {2006},
  doi          = {10.1088/0305-4470/39/43/009}
}

@article{Zhu2011,
  author       = {Huangjun Zhu and Berthold-Georg Englert},
  title        = {Quantum State Tomography with Fully Symmetric Measurements and Product Measurements},
  journal      = {Physical Review A},
  volume       = {84},
  number       = {2},
  pages        = {022327},
  year         = {2011},
  doi          = {10.1103/PhysRevA.84.022327}
}

@article{bent:2015,
  title={Experimental realization of quantum tomography of photonic qudits via symmetric informationally complete positive operator-valued measures},
  author={Qassim, H and Tahir, AA and Sych, D and Leuchs, G and Sánchez-Soto, LL and Karimi, E and Boyd, RW},
  journal={Physical Review X},
  volume={5},
  number={4},
  pages={041006},
  year={2015},
  publisher={APS}
}

@article{qassim:2014,
  title={Limitations to the determination of a Laguerre--Gauss spectrum via projective, phase-flattening measurement},
  author={Qassim, Hammam and Miatto, Filippo M and Torres, Juan P and Padgett, Miles J and Karimi, Ebrahim and Boyd, Robert W},
  journal={Journal of the Optical Society of America B},
  volume={31},
  number={6},
  pages={A20--A23},
  year={2014},
  publisher={OSA}
}

@article{berkhout:2010,
  title={Efficient sorting of orbital angular momentum states of light},
  author={Berkhout, Gregorius CG and Lavery, Martin PJ and Courtial, Johannes and Beijersbergen, Marco W and Padgett, Miles J},
  journal={Physical review letters},
  volume={105},
  number={15},
  pages={153601},
  year={2010},
  publisher={APS}
}

@article{james:2001,
  title={Measurement of qubits},
  author={James, Daniel FV and Kwiat, Paul G and Munro, William J and White, Andrew G},
  journal={Physical Review A},
  volume={64},
  number={5},
  pages={052312},
  year={2001},
  publisher={APS}
}

@article{verbeeck2005plasmon,
  title={Plasmon holographic experiments: theoretical framework},
  author={Verbeeck, J. and van Dyck, D. and Lichte, H. and Potapov, P. and Schattschneider, P.},
  journal={Ultramicroscopy},
  volume={102},
  number={3},
  pages={239--255},
  year={2005},
  publisher={Elsevier}
}

@article{harscher1997interference,
  title={Interference experiments with energy filtered electrons},
  author={Harscher, A. and Lichte, H. and Mayer, J.},
  journal={Ultramicroscopy},
  volume={69},
  number={3},
  pages={201--209},
  year={1997},
  publisher={Elsevier}
}

@article{Saitoh:2013,
  title={Measuring the orbital angular momentum of electron vortex beams using a forked grating},
  author={Saitoh, Koh and Hasegawa, Yuya and Hirakawa, Kazuma and Tanaka, Nobuo and Uchida, Masaya},
  journal={Physical review letters},
  volume={111},
  number={7},
  pages={074801},
  year={2013},
  publisher={APS}
}

@article{potapov2007inelastic,
  title={Inelastic electron holography as a variant of the Feynman thought experiment},
  author={Potapov, P. L. and Verbeeck, J. and Schattschneider, P. and Lichte, H. and van Dyck, D.},
  journal={Ultramicroscopy},
  volume={107},
  number={7},
  pages={559--567},
  year={2007},
  publisher={Elsevier}
}

@article{egerton2007electron,
  title={Electron energy-loss spectroscopy in the TEM},
  author={Egerton, R.},
  journal={Ultramicroscopy},
  volume={107},
  number={6},
  pages={575--586},
  year={2007},
  publisher={Elsevier}
}

@article{larocque2016mapping,
  title={Mapping atomic orbitals with the transmission electron microscope},
  author={Larocque, H. and Bouchard, F. and Grillo, V. and Sit, A. and Frabboni, S. and Dunin-Borkowski, R. E. and Padgett, M. J. and Boyd, R. W. and Karimi, E.},
  journal={Physical Review Letters},
  volume={117},
  number={15},
  pages={154801},
  year={2016},
  publisher={APS}
}

\clearpage
\onecolumngrid
\renewcommand{\figurename}{\textbf{Figure}}
\setcounter{figure}{0} \renewcommand{\thefigure}{\textbf{S{\arabic{figure}}}}
\setcounter{table}{0} \renewcommand{\thetable}{S\arabic{table}}
\setcounter{section}{0} \renewcommand{\thesection}{S\arabic{section}}
\renewcommand{\thesubsection}{S.\arabic{subsection}}
\setcounter{equation}{0} \renewcommand{\theequation}{S\arabic{equation}}
\onecolumngrid
\vspace{1 EM}

\section{Supplementary Information for: ``Quantum Tomography of Inelastic Electron Scattering via Orbital Angular Momentum States"}

\subsection*{Density matrix reconstruction \emph{via} constrained maximum-likelihood estimation}
We employed a modified maximum-likelihood estimation (MLE) algorithm implemented in \textit{Wolfram Mathematica} to reconstruct the quantum state of the inelastically scattered electron beam. The algorithm minimises a total cost function $U$, which comprises four terms that enforce consistency with experimental measurements and incorporate physical constraints:
\begin{eqnarray}
U = U_{\mathrm{OAM}} + \lambda_1 U_\theta + \lambda_2 U_{\mathrm{sym}} + \lambda_3 U_{\mathrm{probe}},
\end{eqnarray}
where $\lambda_i$ are tunable weighting factors that balance the contribution of each term.
The primary fit to the experimental data is enforced through two potentials: \underline{OAM marginal constraint:}
    \[
    U_{\mathrm{OAM}} = \sum_{\ell} \left| \rho_{\ell\ell} - I^{\mathrm{(exp)}}(\ell) \right|^2,
    \]
which ensures agreement with the experimentally-measured OAM-resolved intensity profile, and \underline{Azimuthal angular constraint:}
    \[
    U_{\theta} = \sum_{\theta=0}^{2\pi} \sum_{\ell, \ell'} \left| \rho_{\ell\ell'} e^{i (\ell - \ell') \theta} - I^{\mathrm{(exp)}}(\theta) \right|^2,
    \]
which fits the reconstructed angular probability distribution to the experimentally-measured azimuthal intensity profile.\newline

Two additional constraints are introduced to guide the solution based on physical and experimental prior knowledge:
\emph{Symmetry constraint:}
    \[
    U_{\mathrm{sym}} = \sum_{\ell, \ell'} \left|\, |\rho_{\ell\ell'}| - |\rho_{-\ell,-\ell'}| \,\right|^2,
    \]
which enforces approximate symmetry under $\ell \leftrightarrow -\ell$, as expected from the structured probe and isotropic sample, and \emph{Probe similarity constraint:}
    \[
    U_{\mathrm{probe}} = \sum_{\ell, \ell'} \left|\, |\rho_{\ell\ell'}| - |\rho^{(0)}_{\ell\ell'}| \,\right|^2,
    \]
which ensures that the reconstructed density matrix remains close (in modulus) to the density matrix $\rho^{(0)}$ of the initial probe state. This contraint is applied only to the magnitudes of the matrix elements, since the global and relative phases may vary due to alignment uncertainties in the sorter.
In order to guarantee that the estimated matrix $\hat{\rho}$ remains a valid physical density matrix (\emph{i.e.}, Hermitian, positive semidefinite, and trace-one) at each optimisation step, we express $\hat{\rho}$ in the form:
\begin{align}
\hat{\rho} = \frac{T T^\dagger}{\mathrm{Tr}(T T^\dagger)},
\end{align}
where $T$ is a lower-triangular complex matrix, and $T^\dagger$ denotes its Hermitian conjugate. This Cholesky-like decomposition ensures positive semidefiniteness by construction.

Numerical minimisation of the total potential $U$ was carried out using built-in nonlinear optimisation routines in \textit{Mathematica}, with gradient descent initialised from the elastic state probe density matrix.

The initial probe was modelled as a nearly pure quantum state given by the expression:
\begin{align}
\ket{\psi_0} = N\left( \ket{-4} + \ket{+4} + 0.14 \ket{0} \right),
\end{align}
where $N$ is a normalisation factor. The corresponding density matrix $\rho^{(0)} = \ket{\psi_0}\bra{\psi_0}$ was used both for initialisation and as a reference in the similarity constraint $U_{\mathrm{probe}}$.

\subsection*{Calculation of the nominal density matrix}

The petal beam used in our experiment was generated by using a binary phase hologram to impart an azimuthal modulation to the electron wavefront. Ideally, the phase modulation introduced by the hologram is described by the expression:
\[
\phi(\theta) = \phi_0 + \frac{\pi}{2} \, \mathrm{sign}\left(\sin(4\theta)\right),
\]
where $\theta$ is the azimuthal coordinate and the $\mathrm{sign}(\sin(4\theta))$ function imposes eight-fold rotational symmetry on the phase. This binary phase pattern ideally generates a superposition of OAM eigenstates with $\ell = \pm 4$, forming a petal-shaped transverse intensity distribution with eight lobes. In practice, the phase hologram is realised by modulating the thickness of a supporting membrane, typically using focused ion beam (FIB) milling. This fabrication method introduces undesired absorption effects due to local variations in material thickness. As a result, the transmission amplitude of the hologram is not uniform and can be approximated in the form:
\[
A(\theta) = A_0 + \Delta A \, \mathrm{sign}\left(\sin(4\theta)\right),
\]
where $A_0$ is the average amplitude transmission and $\Delta A$ represents the contrast induced by the modulation.

The total wavefunction at the exit surface of the hologram, when illuminated by a plane wave, is given by the expression:
\[
\psi(\theta) = \left(A_0 + \Delta A \, \mathrm{sign}\left(\sin(4\theta)\right)\right) 
\exp\left(i \frac{\pi}{2} \, \mathrm{sign}\left(\sin(4\theta)\right)\right),
\]
revealing that the beam is a superposition of a pure phase-modulated wave and a weak amplitude-modulated component. The resulting modulation can be interpreted as a convolution of amplitude and phase holographic effects. 

If the Fourier decomposition of the field is restricted to low-order modes ($|\ell| \leq 5$), then the wavefunction can be approximated as:
\[
\psi(\theta) \approx A_0 \, \exp\left(i \frac{\pi}{2} \, \mathrm{sign}(\sin(4\theta))\right) 
+ \Delta A \, \mathrm{sign}(\sin(4\theta)).
\]
This form reveals the principal contribution of a binary phase grating with eight-fold periodicity, while the amplitude term introduces an additional four-fold modulation.

Expanding the phase-modulated component in Fourier space gives:
\[
\psi(\theta) \approx a_0 + 2a_4 \sin(4\theta) 
= a_0 + a_{+4} e^{i4\theta} + a_{-4} e^{-i4\theta},
\]
where $a_{\pm4}$ are complex coefficients corresponding to the $\ell = \pm 4$ OAM modes and $a_0$ is a residual $\ell = 0$ component resulting from the amplitude modulation.

By using this expression, the nominal density matrix of the generated beam can be constructed in the form:
\[
\rho^{(0)} = \ket{\psi}\bra{\psi},
\]
which represents a coherent superposition dominated by $\ell = \pm 4$, with minor contributions from $\ell = 0$ due to the imperfect binary amplitude pattern. In the ideal case ($\Delta A = 0$), $a_0 \rightarrow 0$, and the beam approaches a pure petal state. The full matrix $\rho^{(0)}$ in the OAM basis truncated to $|\ell| \leq 5$ is calculated and used in the reconstruction algorithm as a reference state.

\subsection*{OAM-EELS with partial normalization}
In order to better visualise the redistribution of orbital angular momentum (OAM) as a function of energy loss, the doubly-dispersed OAM-EELS spectrum was analysed by applying a partial normalisation. Specifically, the OAM distribution at each energy loss was normalised to its local maximum. This normalisation removes the global decay of signal intensity due to inelastic cross-section variation and highlights relative changes in the OAM distribution with energy.

\begin{figure}
   \centering
    \includegraphics[width=0.65\columnwidth]{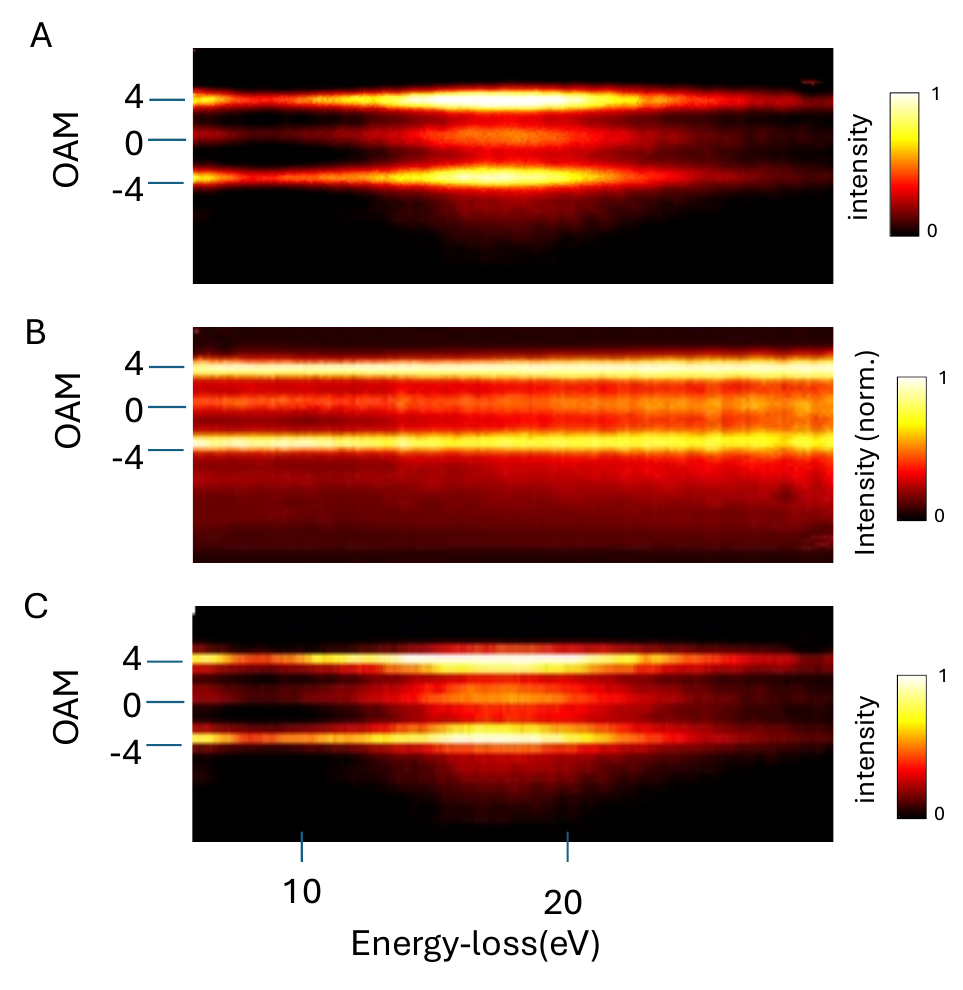}
		\caption{\textbf{(A)} OAM-resolved electron energy-loss spectrum for an input petal beam. In \textbf{(B)}, the OAM spectrum has been normalised to its local maximum at each energy loss. The results highlight the relative redistribution of OAM states across the energy spectrum. As the energy loss increases, the relative population of lower-order OAM components ($|\ell| < 4$) increases compared to the dominant $\ell = \pm 4$ peaks, revealing the onset of angular momentum mixing due to inelastic scattering. In \textbf{(C)}, the OAM spectral data have been discretized for input to the ML algorithm.}
		\label{fig:FigS1}
\end{figure}

The resulting false-colour map (Fig.~\ref{fig:FigS1}) reveals a clear trend. As the energy loss increases, the relative population of OAM states with $|\ell| < 4$ grows compared to the dominant $\ell = \pm 4$ components. This observation is consistent with the expected broadening of the angular momentum distribution due to momentum exchange during plasmon excitation. The method is particularly useful in emphasizing the emergence of off-diagonal and dipolar transitions ($\Delta \ell = \pm 1$), which may be less visible in unnormalized datasets.

\subsection*{Density matrix reconstruction in the elastic case}

For comparison with the inelastic case, we reconstructed the density matrix of the electron beam under elastic conditions using the same maximum-likelihood framework that was described previously. This calculation provides a baseline reference for the degree of coherence and the structure of the initial probe. The reconstructed density matrix for the elastic case closely resembles the ideal model based on the nominal petal beam, with dominant contributions at $\ell = \pm 4$ and a weak residual $\ell = 0$ component (Fig.~S2). The consistency between the reconstructed and theoretical density matrices confirms the effectiveness of the preparation method, while providing a benchmark for assessing coherence loss after inelastic scattering.
Deviations from the idealised model, in particular the slight population of intermediate OAM states and small asymmetries between $\ell = +4$ and $\ell = -4$, can be attributed primarily to limitations of the OAM sorter. A constraint is the practical impossibility of fully separating the radial and angular degrees of freedom. In the current implementation, residual radial mixing, particularly near the needle axis, introduces mode overlap that leads to imperfect mode discrimination. This effect is further compounded by the finite energy and spatial resolution of the spectrometer, as well as by small misalignments in the sorter optics.
Despite these limitations, the elastic case reconstruction serves as a high-coherence reference point for evaluating inelastic scattering processes and validating the density matrix retrieval algorithm.
\begin{figure}
   \centering
    \includegraphics[width=1.0\columnwidth]{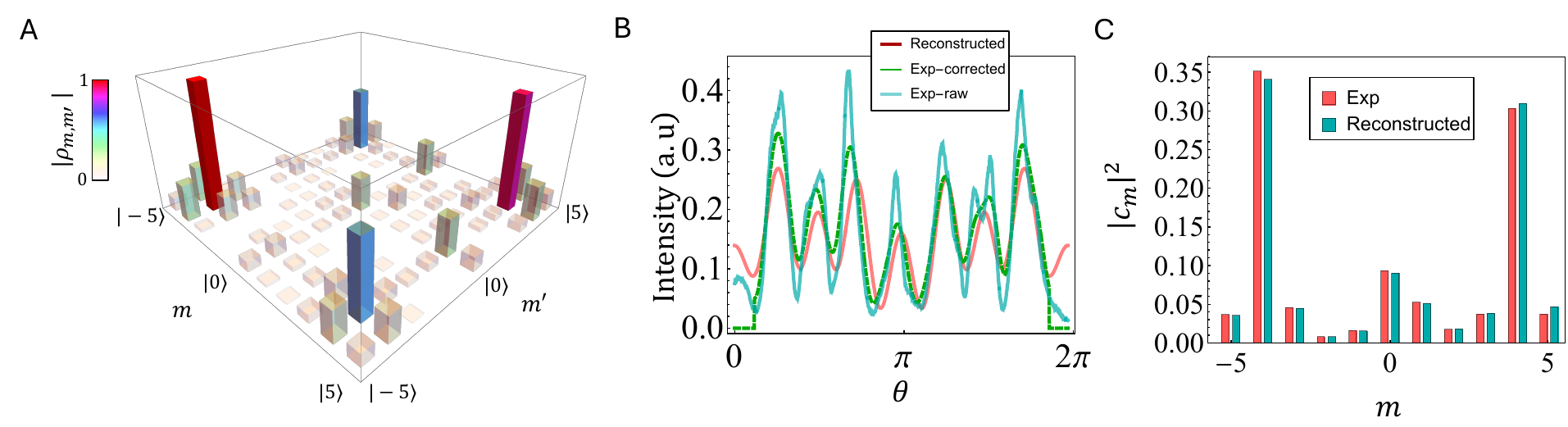}
		\caption{Comparison between the experimentally-retrieved density matrix of the elastic probe and the ideal petal beam model. The differences can be attributed largely to limitations of the OAM sorter, and specifically to its inability to fully decouple the radial and OAM degrees of freedom. This coupling introduces residual mode mixing and prevents perfect discrimination of the OAM eigenstates, thereby contributing to the off-diagonal and intermediate-intensity components observed in the experimental density matrix.} 
		\label{fig:FigS2}
\end{figure}

\subsection*{Density matrix formalism and connection to Fermi's golden rule}

We consider a quantum system composed of two subsystems, $A$ and $B$, initially unentangled and described by a factorised pure state:
\[
\psi = \psi_A \otimes \psi_B,
\]
with the corresponding density matrix:
\[
\rho_0 = |\psi\rangle \langle\psi|.
\]

The subsystems evolve under their respective non-interacting Hamiltonians until they interact briefly through an interaction potential $V_{AB}$, while the full system remains isolated. The final state is given, to first order, by the expression:
\[
|\psi_f\rangle = e^{icV} |\psi_0\rangle \approx (1 + icV)|\psi_0\rangle.
\]

Assuming that only subsystem $A$ (the electron) is accessible experimentally, the reduced density matrix is obtained by tracing over subsystem $B$ (the sample):
\[
\rho_A = \mathrm{Tr}_B(\rho_f).
\]

The interaction potential $V$ can be expressed using a Schmidt (or singular value) decomposition:
\[
V = \sum_{i,j} v_{ij} |i\rangle_A |j\rangle_B \langle i|_A \langle j|_B,
\]
where $\{|i\rangle\}$ and $\{|j\rangle\}$ form orthonormal bases for systems $A$ and $B$, respectively.

For example, in the case of a point-like interaction that conserves total angular momentum, $\braket{\mathbf{k}}{m}_A = f(|k|)e^{i m\theta}$ and $\braket{\mathbf{k}}{n}_B = g(|k|)e^{-i n\theta}$, such that $v_{nm} = \delta_{nm} \tilde{v}(n)$. Neglecting spatial overlap, the reduced density matrix becomes:
\[
\rho_A = \sum_i (1 + c \tilde{v}_i) \rho_{0,ii} |i\rangle \langle i|.
\]

This expression shows that, to first order, the eigenbasis of $\rho_A$ reflects the interaction potential, and its eigenvalues represent transition probabilities. The off-diagonal terms of the initial pure state $\rho_0$ do not contribute, reinforcing the diagonal structure post-interaction.

Therefore, the experimentally reconstructed density matrix $\rho_A$ encodes both the selection rules and the structure of the perturbative interaction. The corresponding states $|j\rangle$ in system $B$ are not accessible without complete knowledge of the interaction potential, but conservation laws (\emph{e.g.}, angular momentum) can allow inference of transitions within the sample.

\subsection*{Plasmon density matrix representation}

The density matrix in transferred momentum coordinates $(\mathbf{q}, \mathbf{q}')$ has the form:
\[
\rho_A(\mathbf{q}, \mathbf{q}') = \langle k_f | e^{-i \mathbf{q}' \cdot \mathbf{r}_e} | k_i \rangle \langle k_i | e^{i \mathbf{q} \cdot \mathbf{r}_e} | k_f \rangle \, G(\mathbf{k}_p) \, \delta(\mathbf{q} - \mathbf{k}_p) \, \delta(\mathbf{q}' - \mathbf{k}_p) \, \delta(\mathbf{q} - \mathbf{q}').
\]

This expression enforces the system to have no coherence between distinct momentum transfers $\mathbf{q} \neq \mathbf{q}'$. The same density matrix can be expressed in the OAM-radial basis after tracing over the radial coordinate in the form:
\[
\rho_{\ell\ell'}(\Delta E) = \int_{|\mathbf{k}| < k_{p,\mathrm{max}}(\Delta E)} \langle \ell | e^{-i \mathbf{q} \cdot \mathbf{r}_e} | \psi_{\mathrm{ei}} \rangle \langle \psi_{\mathrm{ei}} | e^{i \mathbf{q} \cdot \mathbf{r}_e} | \ell' \rangle \, G(\mathbf{k}_p) \, d|\mathbf{k}_p|.
\]

This approach allows angular momentum exchange to be analysed directly using a reduced density matrix over the OAM basis.

\subsection*{Monte Carlo simulation of the density matrix}

We use a Monte Carlo approach to model inelastic scattering in a realistic framework. The interaction is treated as a momentum-space translation of the probe wavefunction due to a momentum transfer $\mathbf{q}$ randomly drawn from a distribution $G(\mathbf{q})$.

For each realization:
\begin{enumerate}
    \item The condenser-plane image of the probe is shifted laterally according to $\mathbf{q}$.
    \item The resulting image is converted into log-polar coordinates: $\theta = \mathrm{atan2}(y, x)$, $\kappa = \log(x^2 + y^2)$.
    \item A 2D Fourier transform is applied to obtain the wavefunction in $(\ell, \kappa)$ space.
    \item The pure-state density matrix is formed and a partial trace over the radial coordinate is performed:
    \[
    \rho_{\ell\ell'} = \sum_\kappa \psi^*(\ell, \kappa) \psi(\ell', \kappa).
    \]
\end{enumerate}

The final OAM density matrix is obtained by averaging over $N$ realizations:
\[
\rho_{\ell\ell'} = \frac{1}{N} \sum_{\mathbf{q}} \rho^{(\mathbf{q})}_{\ell\ell'}.
\]

This method accounts for incoherent averaging over momentum exchanges and enables the extraction of physically meaningful coherence loss in the reconstructed density matrix.
\begin{figure}
   \centering
    \includegraphics[width=0.675\columnwidth]{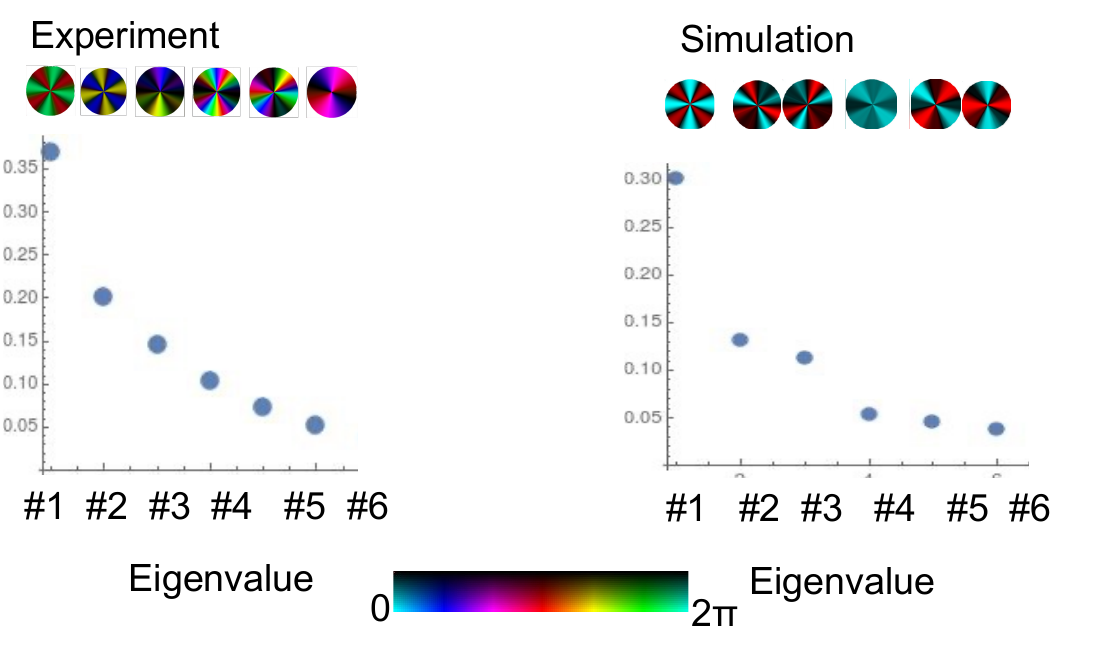}
		\caption{Spectra of the experimental and theoretical density matrix for an input petal beam ($\abs{l}=4$). The plots show the eigenvalues, while the insets show the corresponding eigenstates.} 
		\label{fig:FigS3}
\end{figure}

\begin{figure}
   \centering
    \includegraphics[width=0.425\columnwidth]{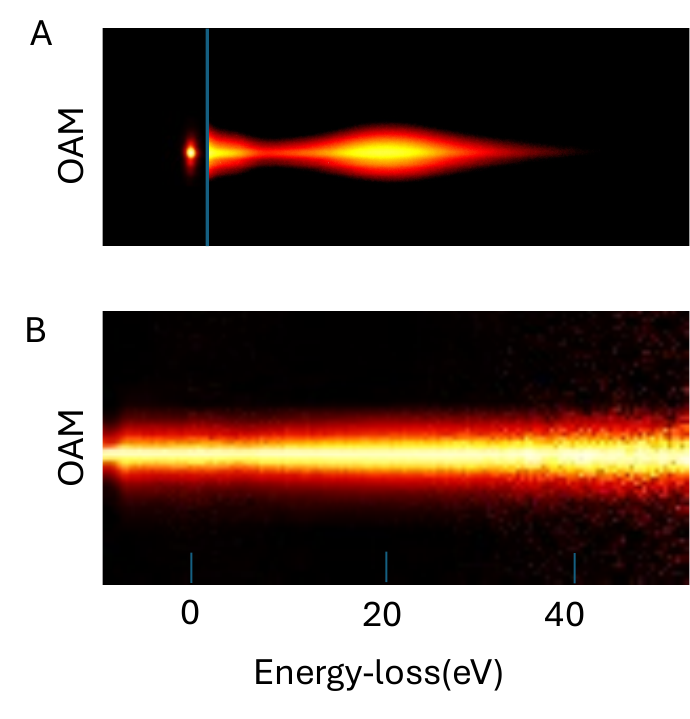}
		\caption{\textbf{(A)} OAM-EELS spectrum for an input beam with zero OAM. \textbf{(B)} The same spectrum with each section of fixed energy loss normalized to its maximum value. The width along the OAM axis increases with energy loss showing coupling with non-zero OAM values due to interaction with the sample.} 
		\label{fig:OAM_0}
\end{figure}

\subsection*{Simulation of chromatic effects}
We aim to evaluate the effect of chromatic aberration in the sorter system. We consider a system composed of two sorter elements and two lenses (the objective lens and a diffracting lens), as shown in Figure S5-A.
\begin{figure}
   \centering
    \includegraphics[width=\textwidth]{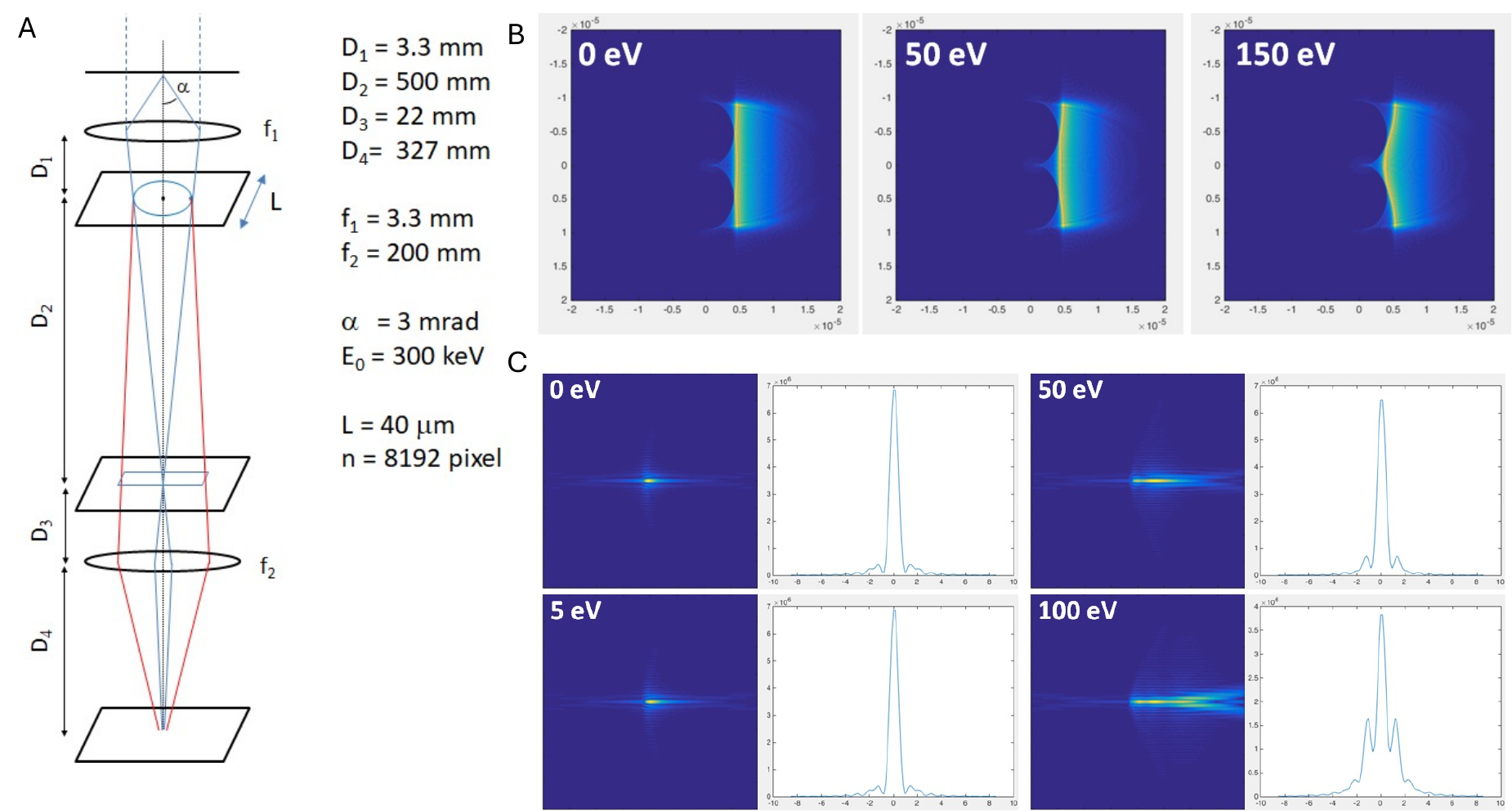}
		\caption{Chromatic effects of sorter elements on the final OAM distribution. \textbf{(A)} Geometry of the experimental apparatus. \textbf{(B)} Wavefunction amplitude distributions at the unwrapper (sorter~2) plane for different electron energies. \textbf{(C)} Final amplitude distributions, with the vertical axis corresponding to the OAM spectrum axis and the horizontal axis to the radial direction. The line profiles show the corresponding distribution along the OAM axis for different energies.}
		\label{fig:chromaticeffects}
\end{figure}
The beam is free-space-propagated between the elements using the Fresnel integral. 
Mathematically, a lens is described by the complex function $\Psi_L=\exp(i ( \pi r^2)/f\lambda)$. In order to correctly compute the effect of the lenses, the phase should not vary too fast. The phase difference between adjacent pixels should be less then $\pi$: $d \phi/d n<\pi$.
Based on this criterion, the minimum focal distance that can be used is $f_{min}=(2L)^2/\lambda n\approx198$ mm.
Therefore, it is not possible to numerically simulate the objective lens, which has a focal distance of 3.3 mm. In order to overcome this limitation, we assume plane wave illumination on the objective lens and a focal distance equal to the distance to sorter~2 ($D=500$ mm).

The focal length of a lens varies linearly with accelerating voltage, according to the expression
$df_1=f_1  dE/E_0 $
A change of defocus in one plane results in a change of defocus in a conjugate plane magnified by a lateral magnification factor $M$, given by $M^2$. This is a simple result of the fact that Fresnel features scale as $df\propto a^2$. Therefore, enlarging a feature a by a factor $M$ brings the same evolution if the defocus is increased by a factor $M^2$.
The defocus $df$ at the sample plane becomes on the sorter~2 plane:
$df_1=f_1  dE/E_0  M^2=dE/E_0   D^2/f_1 $,
where $M$ is the magnification in the sorter~2 plane $M=D/f\approx150$. 
The equivalent total focal length of the objective lens becomes:
$f_o=D+\frac{D^2}{f_1}   \frac{dE}{E_0} =D(1+\frac{D}{f_1}\,\frac{dE}{E_0})$.
The chromatic effect of the diffraction lens is instead:
$f_D=f_2 (1+dE/E_0 )$
Chromatic effects of the sorter’s elements have been accounted for by considering the variation of the coupling factor $C_E$ connecting the integrated potential to the phase: 
$\phi_{S1,2} (\Delta E)=(1+(\Delta C_E)/C_E ) \phi_{S1} (\Delta E=0) $  
This effect is negligible when compared to the chromaticity of the lenses.
As shown by simulations (Fig. S5-C), an energy loss of several tens of eV is required to have an observable effect on the shape of the beam. The final OAM spectrum is only slightly affected by the chromaticity of the system and is negligible below 50~eV, which is the energy window explored in the present work. The primary effect of the defocus is in a direction orthogonal to the axis in which the OAM spectrum is mapped. For a perfectly aligned system, we find that the FWHM is not affected by chromatic effects for energy losses below 50~eV. Although misalignment may induce parasitic higher-order effects, particular attention was paid to achieve precise alignment during experiments. 

\subsection{Specifications of electron microscope and effects of sorter obstruction}

Experiments were performed using a Thermo Fisher Scientific TITAN TEM operated at 300~keV. Energy filtering was carried out using a Gatan GIF Tridiem with a dispersion of 50~meV/pixel.

\begin{figure}
   \centering
    \includegraphics[width=0.5\textwidth]{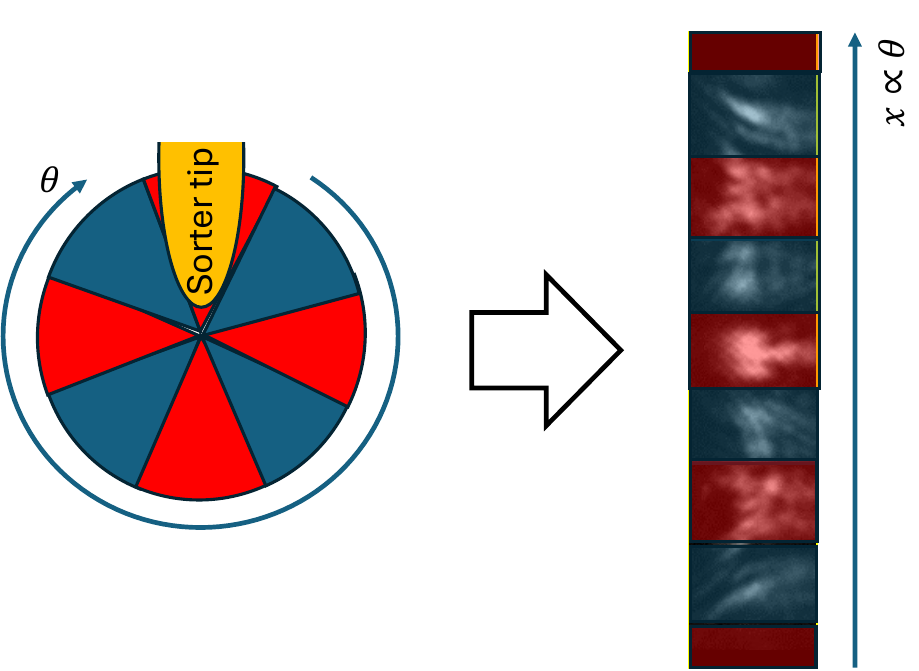}
		\caption{Effect of electrode obstruction on the measured OAM spectrum. }
		\label{fig:obstruction}
\end{figure}

Figure \ref{fig:obstruction} shows the correspondence between a petal beam and an unwrapped beam. 
The presence of sorter electrodes removes part of the intensity, so in effect, only 7 petals are perfectly visible, with a minor disturbance of the Fresnel fringes.

\end{document}